\documentclass[12pt,a4paper]{article}
\usepackage{url}
\usepackage[dvipdfmx]{graphicx}
\usepackage{amsmath,amssymb}
\usepackage{mathrsfs}
\usepackage{comment}
\usepackage{cancel}
\usepackage{xcolor}
\usepackage{cite}
\usepackage[italicdiff]{physics}
\usepackage[colorlinks=true,citecolor=green!60!blue]{hyperref}
\renewcommand\thefootnote{\#\arabic{footnote}}
\newcounter{num}

\begin{document}

\begin{titlepage}

\def\thefootnote{\fnsymbol{footnote}}

\begin{center}

\hfill KANAZAWA-21-05 \\
\hfill January, 2022

\vspace{0.5cm}
       {\Large\bf  Subcritical hybrid inflation \\
       in a generalized superconformal model}

\vspace{1cm}
{\large Yoshihiro Gunji} 
{\large and Koji Ishiwata}

\vspace{1cm}

{\it Institute for Theoretical Physics, Kanazawa University, Kanazawa
  920-1192, Japan}

\vspace{1cm}

\abstract{We study a generalized superconformal model that gives rise
  to a subcritical regime of $D$-term hybrid inflation. Exhibiting the
  model both in a Jordan frame and in the Einstein frame, the
  effective potential of the subcritical regime is derived in the
  Einstein frame. It turns out the inflaton-waterfall field dynamics
  leads to various types of inflaton potential.  Consequently the
  tensor-to-scalar ratio is found to range from $10^{-4}$ ($10^{-3}$)
  to $0.1$ for getting 60 (50) $e$-folds before the end of inflation.

}

\end{center}
\end{titlepage}

\section{Introduction}

Observations of the cosmic microwave background (CMB) radiation
have provided important clues to solving the mysteries of the
Universe. One of the prominent facts revealed by CMB observations
is the indication of inflation at the early stages of the
Universe. Inflation is a paradigm that not only solves the horizon and
flatness problems but also gives the primordial curvature perturbation
leading to the large-scale structure of the present Universe (if dark
matter exists).  Many inflation models have been proposed so far, and
some of them have already been excluded by the CMB observations. The
latest results by the Planck Collaboration on the scalar spectral
index $n_s$, the tensor-to-scalar ratio $r$, and the scalar amplitude
$A_s$ are~\cite{Aghanim:2018eyx,Akrami:2018odb}
\begin{align}
  n_{s}&=0.9649\pm0.0042\,(68\%\,\mathrm{C.L.})\,,
  \label{eq:ns_obs}\\
  r&<0.10\,(95\%\,\mathrm{C.L.})\,,
  \label{eq:r_obs} \\
  A_s&=2.100\pm0.030\times10^{-9}\,(68\%\,\mathrm{C.L.})\,.
  \label{eq:As_obs}
\end{align}
Among the inflation models, the $R^2$ Starobinsky
  model~\cite{Starobinsky:1980te,Mukhanov:1981xt} is a traditional and
  representative model that has good agreement with the
observations. It is also known that similar predictions for $n_s$ and
$r$ are obtained in the $\alpha$ attractor model with small
$\alpha$~\cite{Kallosh:2013yoa}.

Recently $D$-term hybrid inflation in a supersymmetric model has been
revisited, and new features of this model have been unveiled. It was
found that the $\alpha$ attractor appears in the superconformal version of the
model~\cite{Buchmuller:2013zfa}, while a chaotic regime was discovered
in the subcritical regime, where the inflaton field value gets smaller
than the critical-point value of the hybrid
inflation~\cite{Buchmuller:2014rfa,Buchmuller:2014dda}.\footnote{References\,.\cite{Clesse:2010iz,Kodama:2011vs,Clesse:2012dw}
  also point out that inflation lasts below the critical point. In
  the model of the literature, the inflation is induced by a slow-rolling
  waterfall field.} On top of that, the superconformal version of the
model has turned out to have the subcritical regime, and it has both
the feature of the $\alpha$ attractor and natural
inflation~\cite{Freese:1990rb}. Such multiple characteristics in the
subcritical regime of $D$-term hybrid inflation are controlled by
(approximate) symmetries of the K\"{a}hler potential and the
superpotential.  This may relate to the geometry of the
metric. The inflation model under conformal symmetry has recently been
studied in metric-affine geometry instead of Riemannian geometry. It
was shown that the $\alpha$ attractor and natural inflation emerge
depending on the global symmetry imposed on the
model~\cite{Mikura:2020qhc}. Furthermore, $k$-inflation is studied in
the conformal metric-affine geometry~\cite{Mikura:2021ldx}.

The subcritical hybrid inflation has other phenomenological features.
It is free from the cosmic string problem due to the fact that the
inflation continues long enough during the subcritical
regime~\cite{Buchmuller:2014rfa,Buchmuller:2014dda}. Besides, it can
be embedded to the minimal supersymmetric standard model (and its
extension) to give rich phenomenology, such as producing baryon
asymmetry and dark matter and a characteristic pattern of neutrino
masses~\cite{Gunji:2019wtk}. Therefore, it is worth investigating the
subcritical regime in the broad class of the inflation model.

In our paper, we study the subcritical regime of $D$-term hybrid
inflation in a generalized superconformal model. While the model is
formulated in the Einstein frame in the framework of the supergravity model,
we show that it can be formulated in the extended form of the
canonical superconformal supergravity model.  Then we discuss the
dynamics of the inflaton and waterfall fields. It is shown that the
subcritical regime is found in a wide range of parameter space, and that
various forms of the effective potential for inflation are
derived. Consequently, the tensor-to-scalar ratio turns out to be
larger than $\order{10^{-4}\,(10^{-3})}$ for 60 (50) $e$-folds before
the end of inflation. In addition, it is found that the symmetry-enhanced points---namely, the K\"ahler potential without an explicit
superconformal breaking term or integer value for $3\alpha$ that
relates to the dimension of the compactified space in superstring
theory---are consistent with the Planck observation.

The results indicate further possibility to build a more
phenomenologically viable model. One example is the minimal
supersymmetric standard model (MSSM) augmented by the right-handed
neutrinos, which was partly analyzed in Ref.~\cite{Gunji:2019wtk}.  In
this model, one of the right-handed sneutrinos plays the role of an inflaton field, while the right-handed (s)neutrino generates the baryon number
of the Universe. In the analysis of the reference, however, the
neutrino oscillation data are not fully taken into account; meanwhile,
the neutrino sector has been intensively studied by neutrino oscillation
experiments~\cite{Adamson:2013whj,Adamson:2013ue,Abe:2017vif,Abe:2018wpn,Adamson:2017gxd,NOvA:2018gge} and cosmological observations~\cite{Vagnozzi:2017ovm,Aghanim:2018eyx}.
Those data not only constrain such a model but also may give important hints for the new symmetry of the flavor. Non-Abelian discrete symmetry, such as $S_3$, $A_4$, $S_4$, and $A_5$, is one of the viable possibilities in that direction; it has been rigorously studied~\cite{Ma:2001dn,Babu:2002dz,Altarelli:2005yp,Altarelli:2010gt,Ishimori:2010au,King:2013eh,King:2014nza} to explain the mysterious pattern of the neutrino mixing and mass
hierarchy. Thus, the results obtained in the present work would give a
hint towards unveiling the underlying symmetry among inflation, baryogenesis, and the neutrino sector.

This paper is organized as follows: In Sec.~\ref{sec:model}, the $D$-term hybrid inflation in the generalized superconformal model is formulated in a Jordan frame and in the Einstein frame. Dynamics of the inflaton and waterfall fields are discussed in Sec.~\ref{sec:dynamics}, and consequently the effective inflaton potential in the subcritical regime is derived. The cosmological consequences are discussed in Sec.~\ref{sec:cosmology}. Section \ref{sec:conclusion} contains our conclusions and discussion for future work. Throughout this paper, we use the Planck unit, $M_{\rm pl}=1$, unless otherwise stated, and the metric tensor $g_{\mu\nu}$ that gives $\eta_{\mu\nu}={\rm diag}(-1,1,1,1)$ in the flat limit.

\section{The model}
\label{sec:model}
We consider a generalized version of the canonical superconformal supergravity (CSS) model. The CSS model is proposed in Ref.~\cite{Ferrara:2010in}. The model is characterized by two components, the superconformal K\"{a}hler potential ${\cal N}$ and superconformal superpotential ${\cal W}$.  In our paper, we introduce an additional parameter $\alpha\,(>0)$ in the superconformal K\"{a}hler potential:
\begin{align}
  {\cal N} =-|X^0|^2
  \left[1-\frac{|S_+|^2+|S_-|^2+|N|^2}{|X^0|^2}
    -\frac{\chi}{2|X^0|^2}
      \Bigl(\frac{N^2\bar{X^0}}{X^0}
      +\frac{\bar{N}^2X^0}{\bar{X}^0}\Bigr)\right]^\alpha\,,
\end{align}
where we have introduced a real constant
  $\chi$.\footnote{In general, the term proportional to $\chi$ can be more complicated form as shown in Ref.~\cite{Ferrara:2010in}.}
A similar form of ${\cal N}$, but with $\chi=0$, is proposed in the context of the superconformal $\alpha$ attractor~\cite{Kallosh:2013yoa}. Here $X^0$, $S_{\pm}$, and $N$ are chiral superfields that have the local U(1) charges 0, $\pm q$ $(q>0)$, and $0$, respectively.
In our paper, we use the same symbol for a chiral superfield and its scalar field unless otherwise noticed. ($N$ and $S_+$ will be identified as the inflaton and waterfall fields, respectively.)
A nonzero $\chi$ explicitly breaks the superconformal symmetry.  For the superconformal superpotential, on the other hand, we consider a renormalizable Yukawa interaction,
\begin{align}
  {\cal W} = \lambda S_+ S_- N\,,
  \label{eq:W}
\end{align}
where $\lambda$ is a dimensionless constant. Here we ignore possible gauge-invariant and renormalizable terms, such as $S_+S_-X^0$, $N^3$, $X^0N^2$, etc., and stick to the simple model.\footnote{Such extension would be interesting in phenomenological point of view. For example, $N$ can be identified as one of right-handed neutrino that has Majorana mass in Ref.~\cite{Gunji:2019wtk} when $\alpha=1$.} After gauge fixing $X^0=\bar{X}^0=\sqrt{3}$ of the superconformal symmetry, the Lagrangian (in a Jordan frame) is obtained as
\begin{align}
  \frac{{\cal L}_J}{\sqrt{-g_J}} =
  {\cal N}\Bigl(-\frac{1}{6}R_J+{\cal A}^2_\mu \Bigr) 
    -{\cal N}_{\beta \bar{\beta}} g^{\mu\nu}_J
    {\cal D}_\mu z^\beta{\cal D}_\nu \bar{z}^{\bar{\beta}}-V_J\,,
\end{align}
where $R_J$ is the Ricci scalar, ${\cal N}_{\beta \bar{\beta}}\equiv \partial^{2}{\cal N}/\partial z^{\beta}\partial \bar{z}^{\bar{\beta}}$ ($z^{\beta}=S_{\pm},\,N$), $g_{J\mu\nu}$ is the metric tensor, and ${\cal D}_\mu \equiv\partial_\mu -igQA_\mu$ is the covariant derivative. 
Meanwhile, the auxiliary gauge field ${\cal A}_\mu$ defined in Refs.~\cite{Ferrara:2010yw, Ferrara:2010in} has been introduced, and ${\cal A}_\mu$ can be taken to be zero when the scalar part is discussed as described in the literature.  $A_{\mu}$, $g$, and $Q$ are the gauge field, the coupling, and the charge of the U(1) gauge. $V_J=V_J^F+V_J^D$ is the scalar potential in the Jordan frame, where
\begin{align}
  V_J^F&={\cal N}^{\beta \bar{\beta}}
  {\cal W}_{\beta}\overline{{\cal W}}_{\bar{\beta}}\,,
  \\
  V_J^D&=\frac{1}{2}(\Re f)^{ab}{\cal P}_a{\cal P}_b\,.
\end{align}
Here, ${\cal N}^{\beta \bar{\beta}}$ is the $(\beta,\,\bar{\beta})$ component of the inverse of ${\cal N}_{I\bar{J}}$ ($X^I=X^0,\,S_\pm,\,N$), which is defined before gauge fixing; ${\cal W}_\beta \equiv \partial{\cal W}/\partial z^{\beta}$; $f$ is the gauge kinetic function; and ${\cal P}_a = -\eta^\beta_a{\cal N}_\beta-\tilde{\xi}$ (${\cal N}_\beta\equiv \partial {\cal N}/\partial z^\beta$).
$\eta^\beta_a$ is the Killing vector, and in the present case $\eta^\beta_a=Qgz^\beta$ and $f=1$. Since we consider U(1) theory, we omit the index $a$ hereafter.
Note that we have introduced the Fayet-Iliopoulos (FI) term $\tilde{\xi}$ by adopting the procedure in Ref.~\cite{Buchmuller:2012ex}.  According to the literature, an additional term in the Lagrangian is considered:
\begin{align}
  \frac{\Delta {\cal L}_J}{\sqrt{-g_J}}=
  g \frac{-{\cal N}\xi}{3}{\cal P}\,,
\end{align}
where $\xi$ is a constant. We take $\xi>0$ without the loss of generality. This term gives $\tilde{\xi}=g{\cal N}\xi/3$ to get ${\cal P}= -gQz^\beta{\cal N}_\beta-g{\cal N}\xi/3$.

The Lagrangian in the Einstein frame is obtained by the Weyl transformation,
\begin{align}
  g_{J\mu\nu} = \Bigl(-\frac{{\cal N}}{3}\Bigr)^{-1}g_{E\mu\nu}\,,
\end{align}
where $g_{E\mu\nu}$ is the metric in the Einstein frame. Then we obtain
\begin{align}
  \frac{\mathcal{L}_E}{\sqrt{-g_E}} =
  \frac{1}{2}R_E
  -K_{\beta\bar{\beta}}g^{\mu\nu}_E\mathcal{D}_{\mu}
  z^{\beta}\mathcal{D}_{\nu}\bar{z}^{\bar{\beta}}
  -V_E\,,
  \label{eq:scalar-grav_part_of_L}
\end{align}
where $R_E$ is the Ricci scalar in the Einstein frame, and 
\begin{align}
  K&=-3\alpha
  \ln \Bigl(-\frac{\Phi}{3}\Bigr)\,,
  \label{eq:K}\\
  \Phi
  &= -3+|S_{+}|^{2}+|S_{-}|^{2}+|N|^{2}
  +\frac{\chi}{2}(N^{2}+\bar{N}^{2})\,, \\
  V_E&=\Bigl(-\frac{{\cal N}}{3}\Bigr)^{-2}V_J\,,
  \label{eq:V_E}
\end{align}
and $K_{\beta\bar{\beta}}\equiv\partial^{2}K/\partial z^{\beta}\partial
\bar{z}^{\bar{\beta}}$.

Although we have derived it from a Jordan frame, one can derive the Lagrangian in the Einstein frame starting from the K\"ahler potential [Eq.~\eqref{eq:K}] and superpotential [Eq.~\eqref{eq:W}] in the supergravity model.
Then the $F$ and $D$ terms are derived from them as
\begin{align}
    V_E^F &= 
  e^{K}
  (
    K^{\beta\bar{\beta}}D_{\beta}{\cal W}D_{\bar{\beta}}\overline{{\cal W}}
    -3|{\cal W}|^{2}
  )\nonumber\\
  &= 
    \Bigl(-\frac{\Phi}{3}\Bigr)^{1-3\alpha}
  \frac{1}{\alpha}
  \Big[
    \delta^{\beta\bar{\beta}}{\cal W}_{\beta}\overline{{\cal W}}_{\bar{\beta}}
    +
    \frac{1}{\Delta}
    \big|
    \delta^{\beta\bar{\beta}}{\cal W}_{\beta}\Phi_{\bar{\beta}}
    -
    3\alpha {\cal W}
    \big|^{2}
    +
    \frac{9\alpha}{\Phi}(1-\alpha)|{\cal W}|^{2}
  \Big]\,,
  \label{eq:F-term}
  \\
  V_E^D&=
  \frac{1}{2}D^2
  = 
  \frac{g^{2}}{2}
  \big(
    K_{\beta}Qz^{\beta}
    -
    \xi
  \big)^{2}
  \nonumber \\
  &=
  \frac{g^{2}}{2}
  \Big[
    \Bigl(-\frac{\Phi}{3}\Bigr)^{-1}
    \alpha q(|S_{+}|^{2}-|S_{-}|^{2})-\xi
  \Big]^{2}\,.
  \label{eq:D-term}
\end{align}
Here, $K^{\beta\bar{\beta}}$ is the inverse of $K_{\beta\bar{\beta}}$, $D_\beta {\cal W}\equiv{\cal W}_\beta+K_\beta {\cal W}$, ${\cal W}_{\beta}\equiv\partial {\cal W}/\partial z^{\beta}$, $\Phi_{\beta}\equiv\partial \Phi/\partial z^{\beta}$, and $\Delta \equiv \Phi-\delta^{\beta\bar{\beta}}\Phi_{\beta}\Phi_{\bar{\beta}}$.
We have explicitly checked that the sum of Eqs.~\eqref{eq:F-term} and \eqref{eq:D-term} coincides with Eq.~\eqref{eq:V_E}.
We note that when $\alpha=1$, this model reduces to one studied in Refs.~\cite{Buchmuller:2012ex,Buchmuller:2013zfa,Ishiwata:2018dxg}, as expected.

From the scalar potential, the masses of scalar part of the canonically normalized $S_{\pm}$ are given by 
\begin{align}
  m_{\pm}^{2}
  =
  \Bigl(
    -\frac{\Phi}{3}
  \Bigr)^{2-3\alpha}
  \frac{\lambda^{2}}{\alpha^{2}}|N|^{2}
  \mp qg^{2}\xi\,.
\end{align}
As with canonical $D$-term hybrid inflation, $S_{+}$ acquires a tachyonic instability depending on the value of $N$, while $S_{-}$ is stabilized at the origin.
Hereafter, we take $S_-=0$. The critical-point value of $N$ where $S_+$ becomes tachyonic is determined by $m_+=0$.
Since both the real and imaginary parts of $N$ can play the role of inflaton in the model, we take $\phi\equiv\sqrt{2}\mathrm{Re}\,N$ as the inflaton field without the loss of generality.
\footnote{To be explicit, the results for the case where $\Im\,N$ are the inflaton field is obtained by replacing $\chi$ with $-\chi$.  In contrast to the previous studies~\cite{Buchmuller:2014dda,Ishiwata:2018dxg}, we do not restrict our present study to the case of $\chi\simeq -1$ ($+1$) where $\Re\,N$ ($\Im\,N$) has an approximate shift symmetry.  Here, ``approximate'' means that the shift symmetry in the K\"ahler potential is broken by the superpotential. Therefore, the mass of $N$ appears in general [see Eq.~\eqref{eq:mtau}].}
In addition, since the scalar potential depends on $|S_+|$, we define a field $s\equiv\sqrt{2}|S_{+}|$ that we refer to as the waterfall field. Then, defining $\Phi(\phi,s)$ as
\begin{align}
  \Phi(\phi,s)&\equiv
  \Phi|_{\sqrt{2}N=\sqrt{2}\bar{N}=\phi,\,\sqrt{2}|S_{+}|=s,\,S_-=0}
  \nonumber \\
  &= 
  -3+\frac{1}{2}\big(s^{2}+(1+\chi)\phi^{2}\big)\,,
\end{align}
the critical-point value $\phi_c$ should satisfy
\begin{align}
  \Bigl(
    -\frac{\Phi_{c}}{3}
  \Bigr)^{2-3\alpha}
  \phi_{c}^{2}
  &= 
  \frac{2\alpha^{2}}{k}\,,
  \label{eq:critical_point}
\end{align}
where $\Phi_{c}\equiv \Phi(\phi_c,0)$ and 
\begin{align}
  k\equiv \lambda^{2}/qg^{2}\xi\,.
  \label{eq:k}
\end{align}
The number of solutions of Eq.~\eqref{eq:critical_point} depends on the values of $\alpha$ and $\chi$. If it has multiple solutions, the potential gets complicated, and it becomes different from the potential in the canonical hybrid inflation. 
In our study, we focus on the case where there is one critical point. In that case, the valid parameter space is
\footnote{When $\chi=-1$, $\phi^2_c=2\alpha^2/k$ for any value of $\alpha$. This case corresponds to the subcritical hybrid inflation with shift symmetry, which is already studied in Refs.~\cite{Buchmuller:2014rfa,Buchmuller:2014dda}.}
\begin{align}
  \left\{
  \begin{array}{ll}
  (\mathrm{i})&~\chi <-1~ {\rm and}~(0<)\,\alpha \le 1
  \\
  (\mathrm{ii})&~\chi >-1~{\rm  and}~\alpha \ge 2/3
  \label{eq:region2}
  \end{array}
  \right.\,.
\end{align}
In the parameter space, $\alpha=1$ and $2/3$ are special values, since the critical-point values can be obtained analytically as
\begin{align}
  \phi_c =
  \left\{ \begin{array}{ll}
  \sqrt{6/(3k+1+\chi)}
    & ~~{\rm for~}\alpha=1 
    \\[2mm]
  (2/3)\sqrt{2/k}
  & ~~{\rm for~}\alpha=2/3
  \end{array} \right.\,.
\end{align}
It is obvious that $k$ has a bound:
\begin{align}
  k>\left\{
  \begin{array}{ll}
  -(1+\chi)/3  & ~~{\rm for}~\alpha=1~{\rm and}~\chi<-1
  \label{eq:bound_k_alpha1}
  \\[2mm]
  4(1+\chi)/27 & ~~{\rm for}~\alpha=2/3~{\rm and}~\chi>-1
  \end{array} \right. \,.
\end{align}
The latter is given by $\phi_c<\phi_{\rm max}\equiv\sqrt{6/(1+\chi)}$,
where $\phi_{\rm max}$ is determined by
  $\Phi(\phi_{\rm max},0)=0$.

For successful inflation, the stabilization of the imaginary part of $N$, which we define as $\tau \equiv \sqrt{2}\Im N$, should be guaranteed in the subcritical region where $\phi<\phi_c$. The mass of $\tau$ in the region is given by
\begin{align}
  m^2_\tau = \frac{g^2\xi^2 k}{\alpha^2}
   \Bigl(-\frac{\Phi_0}{3}\Bigr)^{1-3\alpha}
   \Bigl(1-\Psi(\phi)\Bigr)
   \Bigl[1-\frac{\phi^2}{6}\bigl\{3-\chi+3\alpha(\chi-1)\bigr\}\Bigr]\,,
     \label{eq:mtau}
\end{align}
where $\Phi_{0}\equiv \Phi(\phi,0)$ and 
\begin{align}
  \Psi(\phi)
  \equiv 
    \Bigl(
      \frac{\Phi_{0}}{\Phi_{c}}
    \Bigr)^{2-3\alpha}
    \frac{\phi^{2}}{\phi_{c}^{2}}
  = 
  \frac{k}{2\alpha^{2}}
  \Bigl(
    -
    \frac{\Phi_{0}}{3}
  \Bigr)^{2-3\alpha}
  \phi^{2}
  \,,
\end{align}
which satisfies $\Psi(\phi_c)=1$.  Here $s^2\ll \phi^2$ has been used, which will be validated in the later discussion. To satisfy $m^2_\tau>0$ in the valid parameter space above, we find the following preferred regions:
\begin{enumerate}
  \item[(i)] $\chi<-1$:
  \begin{align}
    \begin{cases}
      1/3+2/3(1-\chi)<\alpha\le 1\,.\\
      \alpha<1/3+2/3(1-\chi),~
            {\rm depending~on~other~parameters.}
            \label{eq:region1_2}
    \end{cases}
  \end{align}
\item[(ii)] $\chi>-1$:
  \begin{align}
    \begin{cases}
      \chi<1\,.\\
      \chi \gg 1~{\rm and~small}~\alpha ~({\rm but}~\ge 2/3).
      \label{eq:region2_2}
    \end{cases}
  \end{align}
\end{enumerate}
 In our numerical analysis, we will compute the cosmological consequences in the parameter space given in 
 Eq.~\eqref{eq:region2}
 and see consistency with the above regions. We will see that $m^2_\tau>0$ gives a constraint for the $\chi\gtrsim 5$ case. When $\tau$ is stabilized to the origin, the scalar potential after the critical point is given by
\begin{align}
  V_{\mathrm{tot}}(\phi,s)
  &= 
  V_{E}^F+V_{E}^D\nonumber\\
  &=
  \Big(
    -\frac{\Phi(\phi,s)}{3}
  \Big)^{1-3\alpha}
  \frac{\lambda^{2}}{4\alpha}
  \phi^{2}s^{2}
  +
  \frac{g^{2}}{8}
  \bigg[
  \Big(
    -\frac{\Phi(\phi,s)}{3}
  \Big)^{-1}\alpha q s^{2}
  -2\xi
  \bigg]^{2}\,.
  \label{eq:scalar_potential}
  \end{align}

Even if  $\tau$ is stabilized at the origin, it has a quantum fluctuation  during inflation, which might cause another instability. In the model, $-\Phi$ must be positive.
Therefore, this requirement leads to a constraint on the amplitude of $\tau$:
\begin{align}
  \tau^2<\frac{6}{1-\chi}-\frac{1+\chi}{1-\chi}\phi^2\,,
  \label{eq:tau_max}
\end{align}
for $\chi<1$. (There is no constraint when $\chi\ge 1$.) The quantum fluctuation during inflation is estimated as
\begin{align}
  \tau^2\sim \frac{H^4}{m_\tau^2}
  \sim g^2\xi^2\,,
\end{align}
where $H$ is the Hubble parameter. Here we have used Eq.~\eqref{eq:mtau} and $H\sim g^2\xi^2$. Though $m_\tau=0$ at the critical point, there is a mass term that arises at loop level (see the next section). Even in that case, the estimate for the mass is roughly the same except for an extra loop factor. In the allowed parameter space that we will show in the later analysis, this quantity is extremely smaller than unity. Therefore, the
  condition for Eq.~\eqref{eq:tau_max} is always satisfied. One may worry about the isocurvature induced by $\tau$. We note that since $\tau$ has the same order of decay width as $\phi$ has, $\tau$ decays at the time of reheating.
Assuming that $\tau$ dominantly decays to a radiation bath, the isocurvature that $\tau$ produces has no effect on the later thermal history. In addition, the energy density of $\tau$ during inflation is estimated as $m_\tau^2 \tau^2\sim H^4$, which is much smaller than the inflaton energy density.
Then, inflation driven by the $\phi$-$s$ system is not affected by the quantum fluctuation of $\tau$.
Thus, there is no constraint due to the quantum fluctuation of $\tau$.\footnote{We are indebted to Tomo Takahashi for private communication on this issue.}

\section{Dynamics of inflaton and waterfall fields}
\label{sec:dynamics}

We discuss the dynamics of inflaton and waterfall fields around and after the critical point. In the typical hybrid inflation model, inflation ends at the critical point where the waterfall field becomes tachyonic.
In the subcritical hybrid inflation, by contrast, inflation continues after the critical point.
The crucial point here is a suppressed $\lambda$, as pointed out in Ref.~\cite{Buchmuller:2014rfa}.
If $\lambda \ll 1$, therefore, the subcritical regime is expected to emerge in the generalized framework of the superconformal hybrid inflation model.
In this section, we confirm this and give the effective potential for inflation after the critical point.
See Refs.~\cite{Buchmuller:2012ex,Buchmuller:2014dda,Ishiwata:2018dxg} for details.

Before it reaches to the critical point, we assume that the inflaton field slowly rolls down to the critical point while the other scalar fields are initially stabilized at the origin.
Since the tree-level potential is constant, the motion of the inflaton field is driven by the Coleman-Weinberg potential~\cite{Coleman:1973jx,Buchmuller:2014rfa}, which is given by
\begin{align}
  V_{1L}=\frac{g^4q^2\xi^2}{32\pi^2}L(\Psi)\,,
\end{align}
where $L(x)\equiv (x-1)^2\ln (x-1)+(x+1)^2\ln (x+1)-2x^2\ln x-\ln
16$. The velocity of the inflaton field at the
critical point $\dot{\phi}_c$ is given by
\begin{align}
  \dot{\phi}_c=-\frac{1}{3H_c}\pdv{V_{1L}}{\phi}\eval_{\phi=\phi_c}\,,
\end{align}
where $H_c$ is the Hubble parameter at the critical point and the dot signifies a time derivative.
After the critical point, the tachyonic growth of the waterfall field begins.
In order to describe the dynamics, we use the canonically normalized waterfall field.
Taking $\hat{s}_k$ as a Fourier mode of the canonically normalized waterfall field, the equation of motion is given
by~\cite{Buchmuller:2012ex,Ishiwata:2018dxg,Asaka:2001ez}
\footnote{Here, $k$ stands for momentum. Do not confuse this with the dimensionless parameter $k$ given in Eq.~\eqref{eq:k}.}
\begin{align}
  \ddot{\hat{s}}_k
  +\left(k^2e^{-2H_ct}-\frac{9}{4}H_c^2-\hat{d}^3t\right)\hat{s}_k=0\,,
\end{align}
where $\hat{d}$ is obtained as
\begin{align}
  \hat{d}^3=
  \frac{g^2q\xi(2-(1-\alpha)(1+\chi)\phi^2_c)}{\phi_c(-\Phi_c/3)}
  |\dot{\phi}_c|\,.
\end{align}
It is noted that $\hat{d}>0$ and $-\dot{\phi}_c>0$ are satisfied in the parameter space of 
Eq.~\eqref{eq:region2}.
By solving the equation of motion, we get the variance $\expval{\hat{s}^2(t)}$.
After the decoherence time $t_{\rm dec}$ defined in Ref.\,\cite{Buchmuller:2014rfa}, the variance is matched to the classical field, and the time evolutions of the waterfall field and inflaton field are determined by the classical equations of motion.
Namely, we solve the classical equations of motion of $\phi$ and $s$ with a boundary condition $s(t_{\rm dec})=\sqrt{\expval{\hat{s}^2(t_{\rm dec})}/K_{S_+\bar{S}_+,\,c}}$:
\begin{align}
  &\dot{\phi}+\frac{1}{3HK_{N\bar{N}}}\pdv{V_{\rm tot}}{\phi}=0\,,
  \\
  &\dot{s}+\frac{1}{3HK_{S_+\bar{S}_+}}\pdv{V_{\rm tot}}{s}=0\,,
\end{align}
where 
\begin{align}
  &H=\sqrt{V_{\rm tot}(\phi,s)/3}\,,
  \\
  &K_{N\bar{N}}=\alpha
  \frac{1+\frac{1}{6}\chi(1+\chi)\phi^2-\frac{1}{6}s^2}
       {(-\Phi(\phi,s)/3)^2}\,,
       \label{eq:K_NN}
  \\
  &K_{S_+\bar{S}_+}=\alpha\frac{1-\frac{1}{6}(1+\chi)\phi^2}
  {(-\Phi(\phi,s)/3)^2}\,.
\end{align}
Here $K_{S_+\bar{S}_+,\,c}$ is the value at the critical point.

In order to track the inflation dynamics in the $\phi$-$s$ system, we
define the slow-roll parameters by the Hubble parameter,
\begin{align}
  \epsilon_H\equiv -\frac{\dot{H}}{H^2}\,,~~~
  \eta_H\equiv \epsilon_H-\frac{\ddot{H}}{2H\dot{H}}\,.
\end{align}
We have confirmed that inflation continues after crossing the critical point in the parameter space we are interested in.
In addition, it is found that the dynamics after the critical point is effectively described by the single field.
To be explicit, the waterfall field is relaxed to the local minimum $s_{\rm min}$ after up to a few Hubble times. 
During this period, the inflaton field merely moves and stays at the critical point.  The local minimum value is approximately given by
\footnote{ To be precise, the local minimum should be determined numerically by solving $\partial V_{\rm tot}(\phi,s)/\partial s=0$. However, it turns out that Eq.~\eqref{eq:local_minimum} agrees well with the exact solution when $\xi\ll1$, which is the parameter region on which we focus.}
\begin{align}
  s_{\mathrm{min}}^{2}(\phi)
  &= 
  -\frac{\Phi_{0}}{3}
  \frac{2\xi}{q\alpha}
  \Big(
    1-\Psi(\phi)
  \Big)\,.
  \label{eq:local_minimum}
\end{align}
We have numerically checked that mass of $s$ is larger than the Hubble in the subcritical region.
Namely, the same situation in Refs.~\cite{Buchmuller:2014rfa,Buchmuller:2014dda,Ishiwata:2018dxg} is realized.
Putting $s_{\rm min}$ into $V_{\rm tot}(\phi,s)$ and ignoring parametrically unimportant terms suppressed by $\xi$, the effective potential in the subcritical regime is obtained as
\begin{align}
  V(\phi)
  &= 
  g^{2}\xi^{2}
  \Psi(\phi)
  \Big(
    1-
    \frac{1}{2}
    \Psi(\phi)
  \Big)\,.
  \label{eq:inflaton_potential}
\end{align}
We have confirmed that after the waterfall field relaxes to the local minimum, $\epsilon_H$ and $\eta_H$ coincide with $\epsilon(\phi)$ and $\eta(\phi)$ defined in Eq.~\eqref{eq:epsilon_eta_phi}, respectively, derived from the single-field effective potential $V(\phi)$.
In later analysis, we use the effective potential to discuss the cosmological consequences.
We will see in the next section that the typical inflaton field value that is canonically normalized is super-Planckian. However, the predicted tensor-to-scalar ratio can be much smaller than unity.

Careful readers may worry about the effect of the waterfall field on the adiabatic curvature perturbation. 
It is expected to be negligible since, as we will see in the next section, the trajectory of the subcritical inflation is almost straight along the inflaton field. Such trajectory was already studied and analyzed in Ref.~\cite{Buchmuller:2014rfa}.
\footnote{The inflation trajectory below the critical point is also analyzed in Refs~.\cite{Clesse:2010iz,Kodama:2011vs,Clesse:2012dw}. In that case, the trajectory during inflation is almost the waterfall field direction.  }
The effect can be evaluated by calculating a quantity $e^\beta=1+4\eta_\perp^2 H^2/M^2$ given in Ref.~\cite{Achucarro:2010da} (see also Ref.~\cite{Chen:2009zp}.)
Here we follow the notation of Ref.\,\cite{Achucarro:2010da}. (Do not confuse their $\beta$ with the one defined in this paper.) 
A variable $\eta_\perp$ describes the curvature of the trajectory. 
Namely, $\eta_\perp=0$ means that the trajectory is straight. $M$ is the mass of the field perpendicular to the inflaton direction.
The value $e^\beta$ has an impact on the scalar amplitude as
\begin{align}
  A_s \to A_s'=e^\beta A_s\,.
\end{align}
Using the formula given in Ref.~\cite{Achucarro:2010da}, we have found that $\eta_\perp^2\sim \order{10^{-6}}$ around 60 $e$-folds in the valid parameter regions that will be shown in Fig.~\ref{fig:chi_alpha}.
In addition, $H^2/M^2\sim \xi$, which is suppressed as $\order{10^{-4}}$ (see Figs.~\ref{fig:ns-r_l-x_c=-5}, \ref{fig:ns-r_l-x_c=-1.03}, and
\ref{fig:nsr_lx_a=1_2ov3}).
Consequently, $e^\beta-1\sim \order{10^{-10}}$, which is sufficiently small. Therefore, the effective description in the subcritical regime is valid.

\section{Cosmological consequences}
\label{sec:cosmology}

Now we are ready to discuss cosmological consequences of this model.
We compute the scalar spectral index, the tensor-to-scalar ratio, and
the scalar amplitude and compare them with the latest observational
results [Eqs.~\eqref{eq:ns_obs}--\eqref{eq:As_obs}].

\subsection{Cosmological parameters and overview of the results}

The slow-roll parameters  defined by the effective potential are 
\begin{align}
  \epsilon(\phi)
  \equiv
  \frac{1}{2}
  \Big(
    \frac{V^{\prime}}{V}
  \Big)^{2}\,,\quad
  \eta(\phi)
  \equiv
  \frac{V^{\prime\prime}}{V}\,.
  \label{eq:epsilon_eta_phi}
\end{align}
Here, $V^{\prime}=dV/d\hat{\phi}$ and $V^{\prime\prime}=d^{2}V/d\hat{\phi}^{2}$, and $\hat{\phi}$ is the canonically normalized inflaton field, which is defined by
\begin{align}
  \frac{d\phi}{d\hat{\phi}}
  =
  K_{N\bar{N}}^{-1/2}\eval_{s=s_{\rm min}}\,.
  \label{eq:dphiovdhatphi}
\end{align}
Since the field value of the waterfall field is found to be parametrically much smaller than the inflaton field value, it can be approximated as $s_{\mathrm{min}}\simeq0$ during inflation. Therefore,
\begin{align}
  K_{N\bar{N}} \simeq \frac{3\alpha}{-\Phi_0}
  \Bigl[1+\frac{(1+\chi)^2\phi^2}{-2\Phi_0}\Bigr]\,.
  \label{eq:K_NN_appr}
\end{align}
Inflation ends at
\begin{align}
  \phi=\phi_{\rm end}\equiv{\rm Max}\{\phi_{\epsilon},\phi_{\eta}\}\,.
\end{align}
$\phi_{\epsilon}$ and $\phi_{\eta}$ are determined by $\epsilon(\phi_{\epsilon})=1$ and $|\eta(\phi_{\eta})|=1$, respectively. $n_s$, $r$, and $A_s$ are  determined by the slow-roll parameters as
\begin{align}
  n_{s}
  &=
  1+2\eta(\phi_{*})-6\epsilon(\phi_{*})\,,\\
  r
 & =
  16\epsilon(\phi_{*}) \,, \\
    A_{s}
  &=
  \frac{V(\phi_{*})}{24\pi^{2}\epsilon(\phi_{*})}\,.
\end{align}
Here, $\phi_{*}$ is determined by the number of $e$-folds before the end
of inflation:
\begin{align}
  N_{*}
  =
  \int_{\hat{\phi}_{\rm end}}^{\hat{\phi}_{*}}d\hat{\phi} \frac{V}{V^{\prime}}
  =
  \int_{\phi_{\rm end}}^{\phi_{*}}d\phi
  \sqrt{\frac{K_{N\bar{N}}}{2\epsilon(\phi)}}
  \,,
  \label{eq:N*}
\end{align}
where $\hat{\phi}_*$ and $\hat{\phi}_{\rm end}$ are canonically normalized field values corresponding to $\phi_*$ and $\phi_{\rm end}$, respectively.  In our numerical analysis, we take $q=g=1$ without the loss of generality.  This is due to the fact that when $\xi\ll 1$, which is the case we are interested in, $q$ and $g$ can be absorbed into $\lambda$ and $\xi$ by redefining these as $\bar{\lambda}\equiv \lambda/\sqrt{qg}$ and $\bar{\xi}\equiv g\xi$, respectively. This is easily seen in Eq.~\eqref{eq:inflaton_potential}. Since $\xi$ is cancelled in the slow-roll parameters, $\phi_{\rm end}$ and $\phi_*$ are determined for $\alpha,\, \chi,\, k$, and $N_*$, and they give $n_s$ and $r$.
The median value of the observed scalar amplitude determines the value of $\lambda$ and $\xi$ (also using the value of $k$).

\begin{figure}[tb]
  \begin{minipage}[t]{0.5\hsize}
  \centering
  \includegraphics[keepaspectratio,scale=0.66]{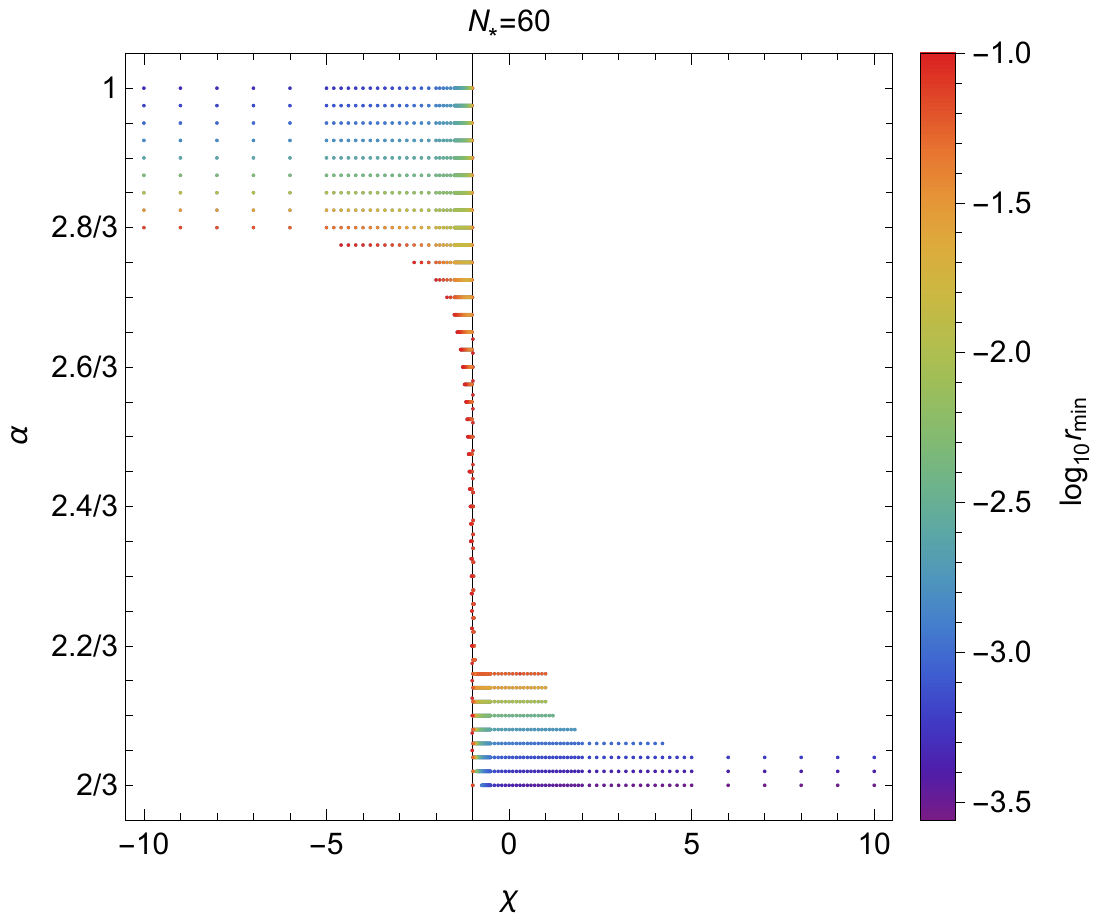}
  \end{minipage}
  \begin{minipage}[t]{0.5\hsize}
  \centering
  \includegraphics[keepaspectratio,scale=0.66]{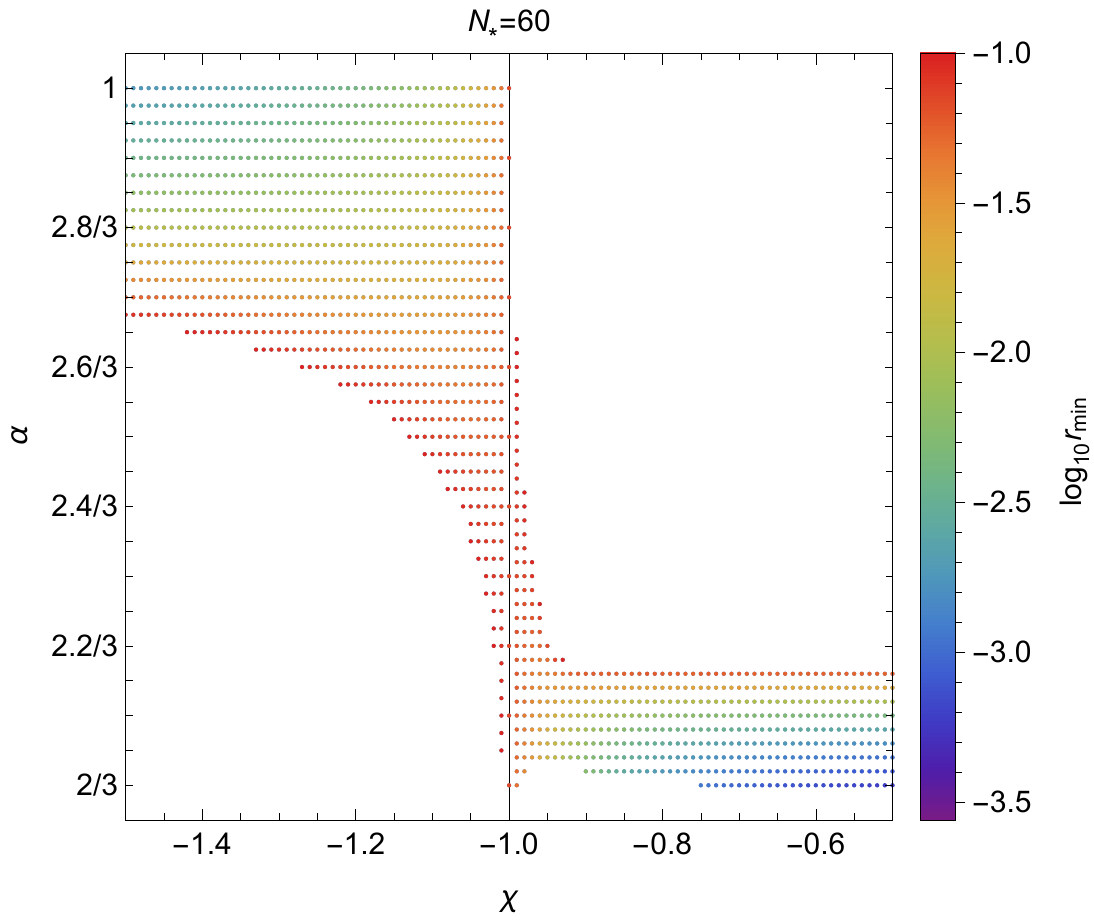}
  \end{minipage}
  \begin{minipage}[t]{0.5\hsize}
  \centering
  \includegraphics[keepaspectratio,scale=0.66]{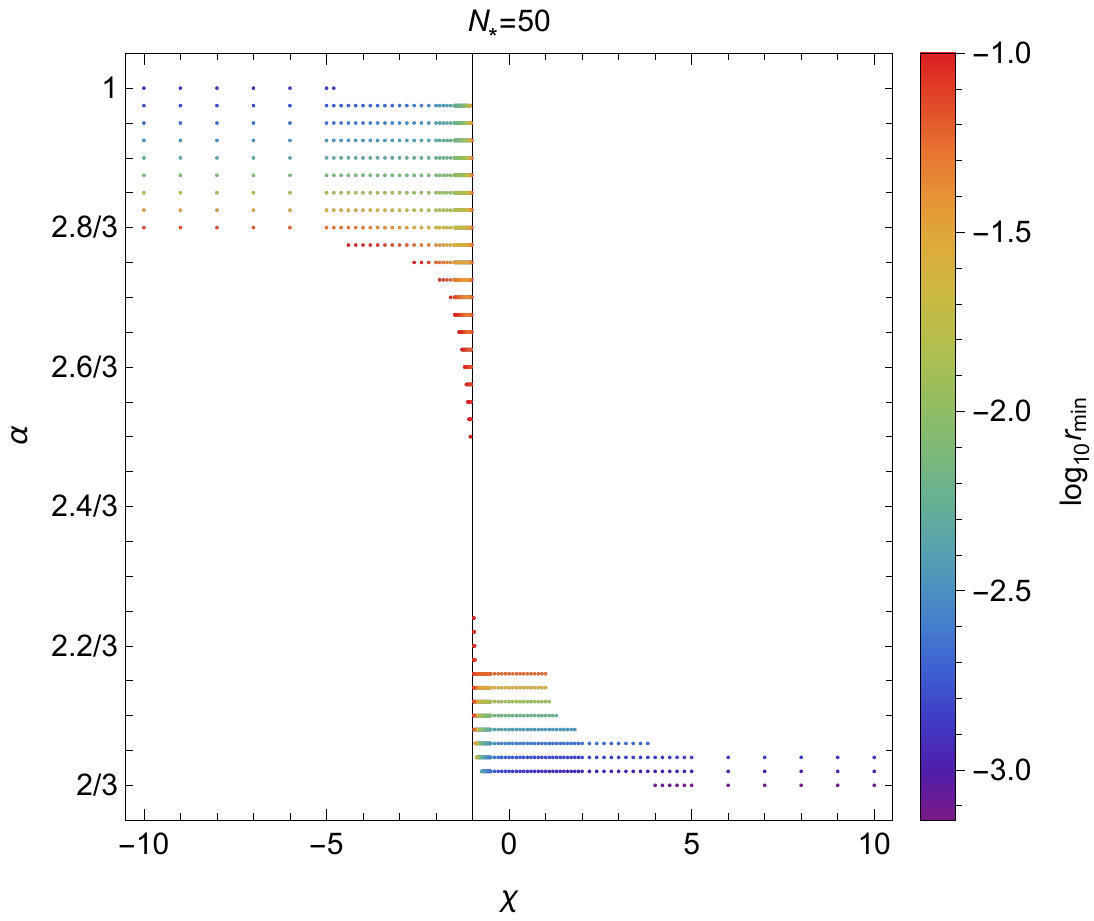}
  \end{minipage}
  \begin{minipage}[t]{0.5\hsize}
  \centering
  \includegraphics[keepaspectratio,scale=0.66]{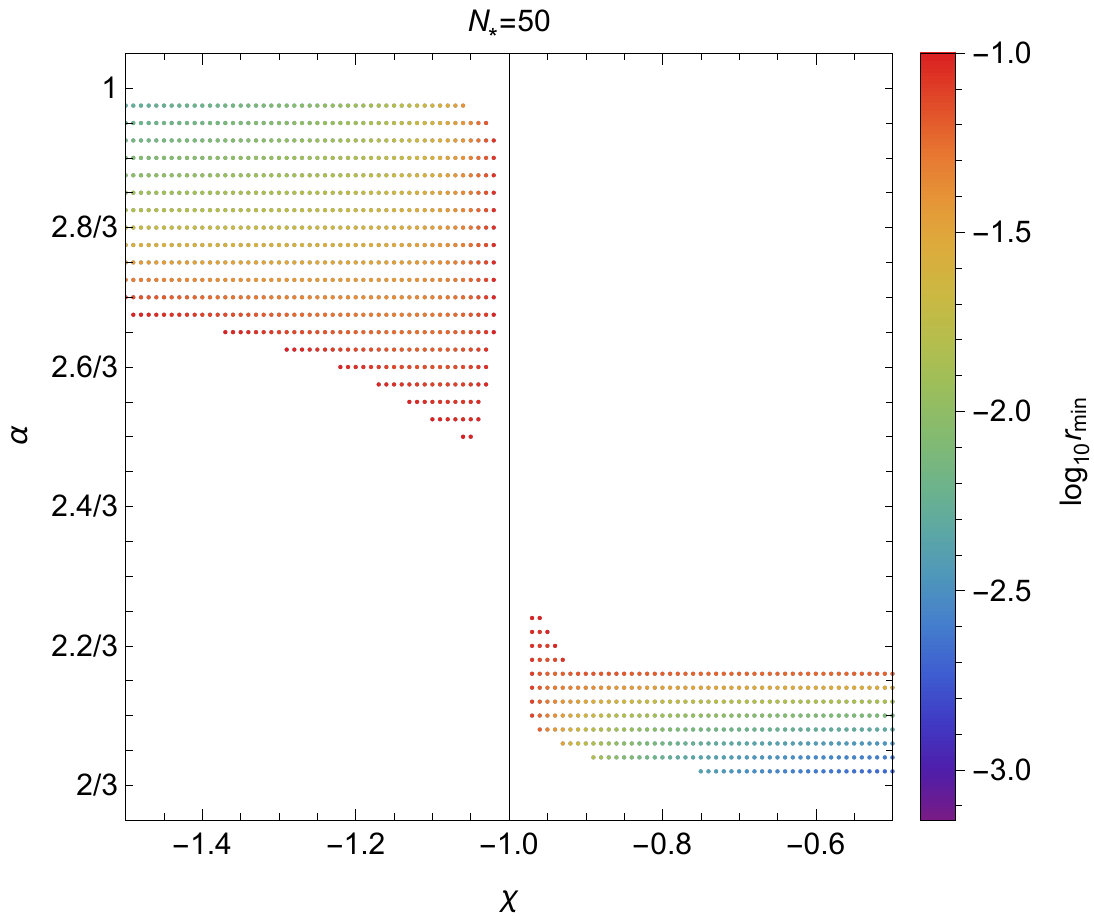}
\end{minipage}
\caption{Allowed region for $\alpha$ and $\chi$ for $N_{*}=60$ (top) and $50$ (bottom) by imposing $m^2_\tau>0$ and Eqs.~\eqref{eq:ns_obs}--\eqref{eq:As_obs}. The right panels are the same as the left ones but the range of $-1.5\le\chi\le-0.5$ is magnified. The color map indicates the minimum value of the predicted tensor-to-scalar ratio, $r_{\mathrm{min}}$.  }
  \label{fig:chi_alpha}
\end{figure}

First of all, we give the allowed region on the $(\alpha,\,\chi)$ plane in Fig.~\ref{fig:chi_alpha}. We impose $m^2_\tau>0$ and the observed results given in Eqs.~\eqref{eq:ns_obs}--\eqref{eq:As_obs}.
In the plot $N_*=60$ and $50$ are taken, and each dot corresponds to an
allowed point.
For a given set of $\alpha$, $\chi$, and $N_*$, the values of $n_s$ and $r$ are given as a function of $k$.
When $k$ has solutions such that the predicted $n_s$ and $r$ are within the range of Eqs.~\eqref{eq:ns_obs} and \eqref{eq:r_obs}, a dot is plotted on the $\alpha$-$\chi$ plane.
The color map shows the minimum value of $r$ in the range of Eqs.~\eqref{eq:ns_obs} and \eqref{eq:r_obs}.
It is found that selective regions are allowed, and the behavior changes around $\chi=-1$.
In addition, the allowed value of $\alpha$ saturates for $|\chi|\gtrsim 5$. When $N_*=60$, for instance, the allowed regions are $\alpha\simeq 1$ for $\chi\lesssim -5$, $2/3 \le\alpha \le 1$ for $\chi\simeq -1$, and $\alpha\simeq 2/3$ for $\chi\gtrsim 5$, and the predicted $r$ changes by orders of magnitude.
It is found that 
Eq.~\eqref{eq:region1_2} does not give constraints on the parameters for $\chi<-1$.
On the contrary, the obtained allowed region roughly tracks the region indicated in 
Eq.~\eqref{eq:region2_2}
for $\chi>-1$.
To investigate further, we categorize the parameter space into two regions:
\begin{align*}
\begin{cases}
  |\chi|\gtrsim 5\\
  \chi\simeq-1
\end{cases}
.
\end{align*}
We examine the dynamics of the inflaton and waterfall fields and their
consequences in detail in Secs.~\ref{sec:caseI} and \ref{sec:caseII}.
In addition, we further investigate the specific cases $\alpha=1$, $2/3$, and $\chi=0$ in Sec.~\ref{sec:caseIII}.
These cases are motivated by theoretical models beyond supergravity. In superstring theory, $3\alpha$ corresponds to the dimension of the compactified space.  Therefore, it is supposed to be an integer. When $\chi=0$, on the other hand, the superconformal invariance is exact in the K\"{a}hler potential, and it corresponds to a symmetry enhanced point.

\subsection{$|\chi|\gtrsim 5$}
\label{sec:caseI}

\begin{figure}[tbp]
    \begin{minipage}[t]{0.5\hsize}
  \centering
  \includegraphics[keepaspectratio,scale=0.8]{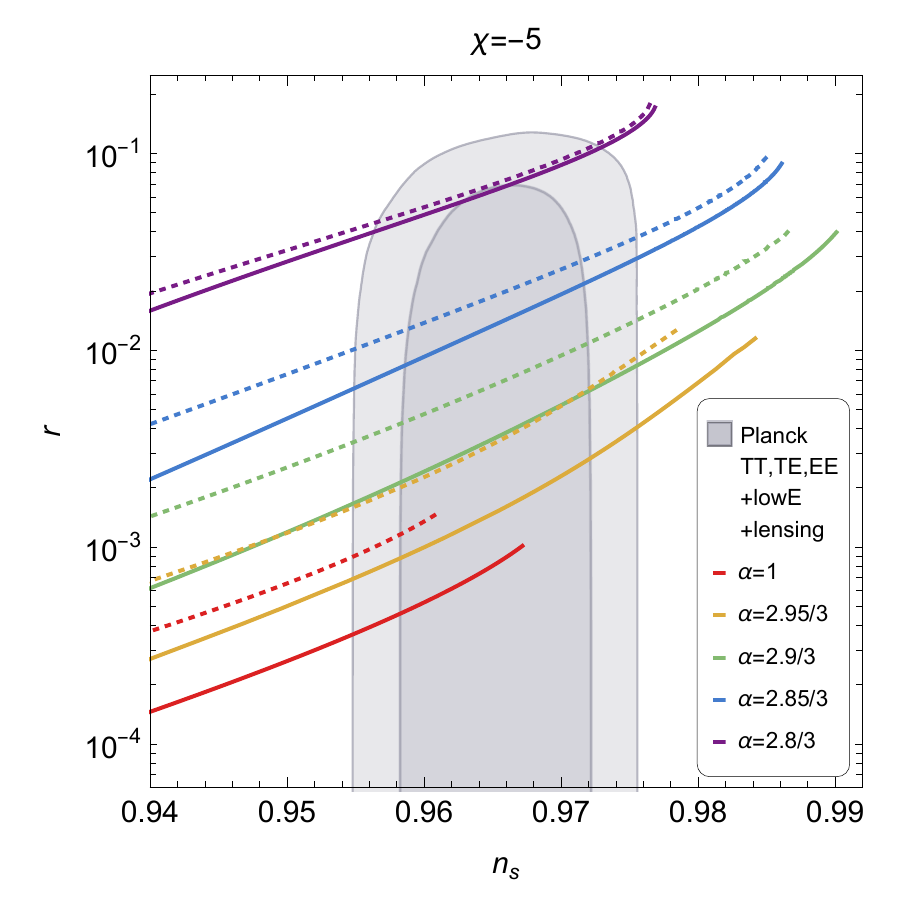}
  \end{minipage}
  \begin{minipage}[t]{0.5\hsize}
  \centering
  \includegraphics[keepaspectratio,scale=0.8]{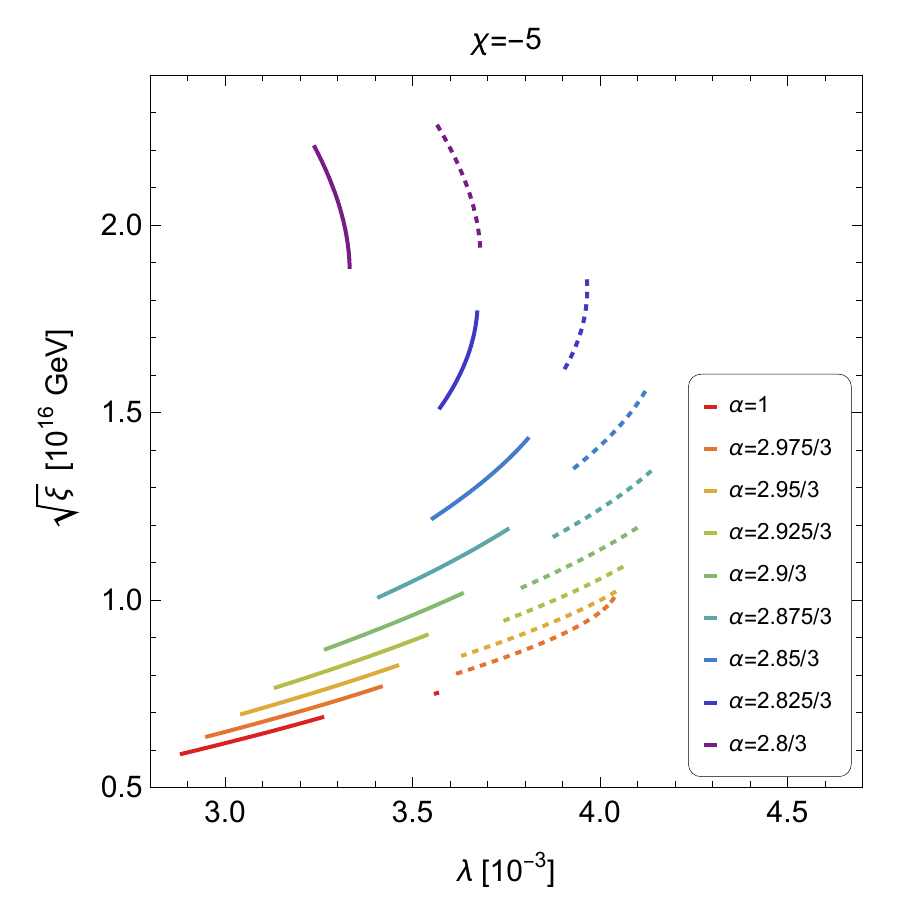}
\end{minipage}
    \begin{minipage}[t]{0.5\hsize}
  \centering
  \includegraphics[keepaspectratio,scale=0.8]{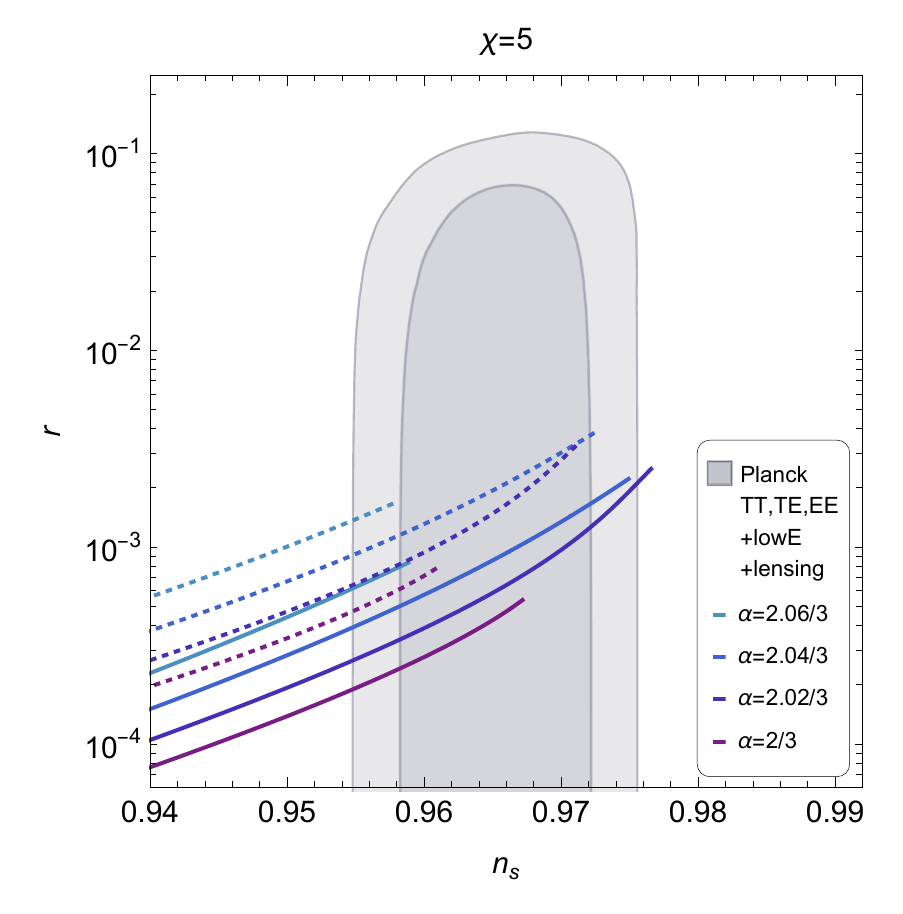}
  \end{minipage}
  \begin{minipage}[t]{0.5\hsize}
  \centering
  \includegraphics[keepaspectratio,scale=0.8]{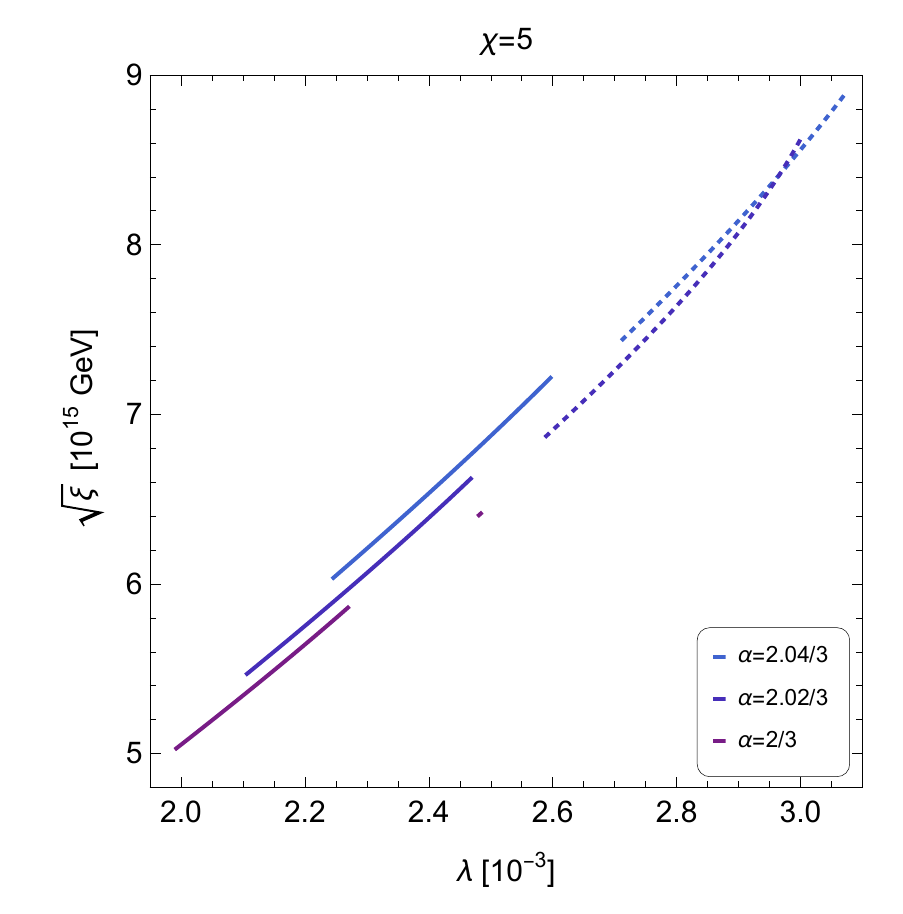}
\end{minipage}
\caption{\textit{Left:} Predicted scalar spectral index and tensor-to-scalar ratiofor fixed $e$-folds, $N_{*}=60$ (solid curves) and $50$ (dotted curves), and various values of $\alpha$.
In the top and bottom panels, $\chi$ is taken to $-5$ and $5$, respectively.
We impose $m^2_\tau>0$, and 1$\sigma$ (dark shaded) and 2$\sigma$ (light shaded) regions from the Planck Collaboration~\cite{Akrami:2018odb} are also shown.
\textit{Right:} Allowed regions for $\lambda$ and $\sqrt{\xi}$ for $\chi=-5$ (top) and $5$ (bottom), by imposing Eqs.~\eqref{eq:ns_obs}--\eqref{eq:As_obs}. }
\label{fig:ns-r_l-x_c=-5}
\end{figure}

\textit{Summary of the predictions.} --- The predictions for $n_{s}$ and $r$ are shown in the left panels of Fig.~\ref{fig:ns-r_l-x_c=-5} for
$\chi=-5$ (top) and $5$ (bottom) with various values of $\alpha$.
Here the stability condition for $\tau$ is imposed.
In the right panels, the parameters ($\sqrt{\xi}$, $\lambda$) that are consistent with the Planck observations are plotted for $N_*=60$ and $50$. For a given $\alpha$, it is found that $n_s$ and $r$ get larger for smaller $k$. For $\chi=-5$, $r$ tends to be smaller when $\alpha \to 1$; meanwhile, it can be as large as $\order{0.1}$. This is the opposite behavior compared to the $\alpha$ attractor model~\cite{Kallosh:2013yoa}.
As a consequence, a lower bound is obtained as $r\gtrsim 10^{-3}$.
For $\chi=5$, on the other hand, $r$ is found to be suppressed as $10^{-4}\lesssim r\lesssim 10^{-3}$, and it gets smaller as $\alpha$ decreases. 
The parameter space that is consistent with the Planck data turns out to be $10^{-3}\lesssim\lambda \lesssim 10^{-2}$ and
$\sqrt{\xi}\sim\order{10^{16}\,{\rm GeV}}$.
The results for $\chi<-5$ ($\chi>5$) behave almost the same as those for $\chi=-5$ ($\chi=5$).
\footnote{We find that the preferred value of $\lambda$ becomes larger, but it is less than $10^{-2}$.}

To get a better understanding of the results, it is legitimate to describe the effective potential in terms of the canonically normalized field $\hat{\phi}$.
In addition, we consider a large field value limit for $\phi$ to give analytical expressions for the effective potential.  Even though the analytical expressions are not always valid, they are used to help understand the numerical results for $n_s$ and $r$ qualitatively.
When $|\chi|\gg 1$ and in a large field limit, the second term in the parenthesis in Eq.\,\eqref{eq:K_NN_appr} can be neglected. 
In that case, the K\"{a}hler metric is approximately given by
\begin{align}
  K_{N\bar{N}}
  \simeq
  \frac{3\alpha(1+\chi)^{2}\phi^{2}}{2\Phi^{2}_0}
  \,,
\end{align}
and consequently Eq.~\eqref{eq:dphiovdhatphi} can be solved analytically as to give $\phi$. Using the result, $\Phi_0$ and $\Psi$ are determined. The results are 
\begin{align}
  \phi^{2}&\simeq
  \begin{cases}
    \displaystyle
  \frac{1}{\beta}
  (Ce^{\sqrt{\frac{2}{3\alpha}}\hat{\phi}}-1)
  &~~~ (\chi<-1)\\[3mm]
    \displaystyle
  \frac{1}{-\beta}
  (1-Ce^{-\sqrt{\frac{2}{3\alpha}}\hat{\phi}})
  &~~~ (\chi>-1)
  \end{cases}\,,
  \label{eq:phi2}
  \\[2mm]
  -\frac{\Phi_0}{3}&\simeq
  \begin{cases}
    \displaystyle
  Ce^{\sqrt{\frac{2}{3\alpha}}\hat{\phi}}
    \displaystyle
  &~~~ (\chi<-1)\\[2mm]
  Ce^{-\sqrt{\frac{2}{3\alpha}}\hat{\phi}}
  &~~~ (\chi>-1)
  \end{cases}
  \,, \\
  \Psi&\simeq
  \begin{cases}
    \displaystyle
  \frac{k}{2\alpha^{2}\beta}
  C^{2-3\alpha}e^{\sqrt{\frac{2}{3\alpha}}(2-3\alpha)\hat{\phi}}
  (Ce^{\sqrt{\frac{2}{3\alpha}}\hat{\phi}}-1)
  &~~~ (\chi<-1)\\[3mm]
    \displaystyle
  \frac{k}{-2\alpha^{2}\beta}C^{2-3\alpha}
  e^{-\sqrt{\frac{2}{3\alpha}}(2-3\alpha)\hat{\phi}}
  (1-Ce^{-\sqrt{\frac{2}{3\alpha}}\hat{\phi}})
  &~~~ (\chi>-1)
  \end{cases}\,,
\end{align}
where $\beta\equiv-(1+\chi)/6$ and $C$ is a positive constant.
$C$ can be determined by, for example, $\Psi(\hat{\phi}_c)=1$. The results are summarized in Table~\ref{table:PsiI}.
For the cases of $\chi< -1$\,\&\,$\alpha\neq 1$ and $\chi> -1$\,\&\,$\alpha\neq 2/3$, $2\alpha^2|\beta|/k\gg 1$ is further assumed, which will be discussed soon.  The approximated expressions for $\Psi$ have similarity even for $\chi<-1$ and $\chi>-1$. Based on the expressions, we categorize the parameter space into two cases:
\begin{align*}
  \left\{
  \begin{array}{l}
    \chi< -1\,\& \,\alpha\neq 1\,\text{or}\,\chi> -1\,\& \,\alpha\neq 2/3\\
    \chi< -1\,\& \,\alpha= 1\,\text{or}\,\chi> -1\,\& \,\alpha= 2/3
  \end{array}
  \right.\,,
\end{align*}
and we discuss the numerical results in Fig.\,\ref{fig:ns-r_l-x_c=-5}.

\begin{table}[t]
 \begin{center}
  \caption{\small Approximated expression of $\Psi$ in the limit $|(1+\chi)^2\phi^2/(-2\Phi_0)|\gg 1$. For cases of $\chi< -1$\,\&\,$\alpha\neq 1$ and $\chi> -1$\,\&\,$\alpha\neq 2/3$, $2\alpha^2|\beta|/k\gg 1$ is further assumed. The parameter space where the approximated expression is valid is given in the right column. (See also Fig.~\ref{fig:chi_alpha} for the allowed region for $\alpha$ and $\chi$.) The effective potential can be obtained from $\Psi$ by using Eq.~\eqref{eq:inflaton_potential}.}
\begin{tabular}{ccc}
  \hline \hline
  Cases   &  Approximated expression for $\Psi$
  & Valid parameters
  \\
  \hline
  $\chi< -1\,$\&\,$\alpha\neq 1$
  & $e^{\sqrt{\frac{6}{\alpha}}(1-\alpha)(\hat{\phi}-\hat{\phi}_c)}$
  & $\chi\lesssim -20$\,\&\,$\alpha\lesssim 2.8/3$\\
  $\chi> -1$\,\&\,$\alpha\neq 2/3$
  & $e^{\sqrt{\frac{2}{3\alpha}}(3\alpha-2)(\hat{\phi}-\hat{\phi}_c)}$
  &$\chi\gtrsim -0.5$\,\&\,$\alpha\gtrsim 2.2/3$
  \vspace{3mm}
  \\
  \begin{tabular}{c}
    $\chi< -1$\,\&\,$\alpha= 1$ \\
    $\chi> -1$\,\&\,$\alpha= 2/3$
  \end{tabular}
 & $ \frac{k}{2\alpha^2|\beta|}
 \left[1-\Bigl(1-\frac{2\alpha^2|\beta|}{k}\Bigr)
   e^{-\sqrt{\frac{2}{3\alpha}}(\hat{\phi}-\hat{\phi}_c)}
   \right]$
 &
 \begin{tabular}{c}
    $\chi\lesssim -10$ \\
    $\chi\gtrsim 2$
  \end{tabular}
  \\
  \hline \hline
 \end{tabular}
 \label{table:PsiI}
\end{center}
\end{table}

\subsubsection{$\chi< -1$\,\&\,$\alpha\neq 1$ or $\chi> -1$\,\&\,$\alpha\neq 2/3$}
As mentioned above, we further assume
  \begin{align}
    2\alpha^2|\beta|/k \gg 1
    \label{eq:condforapp}
  \end{align}
to derive the expressions, which are shown in the first row of Table~\ref{table:PsiI}.
If the above condition is satisfied, then the second term in the parenthesis of Eq.~\eqref{eq:phi2} can be ignored.
(The condition can be rewritten as $|\beta|\phi_c^2\gg 1$.)
We note here that this approximation is only valid for an $\alpha$ which is not much closer to 1 or 2/3.
Ignoring this term under the limit of $\alpha \to 1$ or 2/3, one cannot determine a consistent $C$ from the analytic expressions.
  
To get a very rough picture, let us make more simplifications.
If we further assume that the effective potential is determined by $V\sim g^2\xi^2 \Psi$, then $n_s$ and $r$ are given by $(n_s\,,r)\sim (1-p^2,\,8p^2)$, where $p\equiv \sqrt{6/\alpha}(1-\alpha)$ for $\chi< -1$ and $p\equiv \sqrt{2/3\alpha}(3\alpha-2)$ for $\chi> -1$.
The expressions indicate that for $\chi< -1$ ($\chi> -1$), $n_s$ and $r$ get larger and smaller, respectively, when $\alpha$ approaches unity (2/3).
This estimation is consistent with the results shown in Fig.\,\ref{fig:ns-r_l-x_c=-5}. Although the qualitative behavior can be understood from this rough estimation, quantitative discussion is found to be more complicated.

For the $\chi<-1$ case, the approximation is found to be good for $\chi\lesssim -20$ and $\alpha\lesssim 2.8/3$.
If $\alpha$ becomes larger, we find that the approximated expression of $\Psi$ does not give the correct values of $n_s$ and $r$ even though Eq.~\eqref{eq:condforapp} is satisfied. 
The maximum values of $n_s$ and $r$ come from saturated values of $\phi_*$ in the limit $k\to 0$.

For the $\chi>-1$ case, on the other hand, the expression is found to be valid for $\chi\gtrsim -0.5$ and $\alpha\gtrsim 2.2/3$. However, we have found that the upper bound for $n_s$ (and $r$) is given by the stability of $\tau$.  This is expected from Eq.~\eqref{eq:region2_2}. Therefore, it is difficult to understand the predicted value of $n_s$ and $r$ based on the simple approximation.

Therefore, to see the $\alpha$ dependence on $n_s$ and $r$ for cases of $\chi<-1$ and $\chi>-1$, it is more intuitive to plot the effective potential numerically.
Figure \ref{fig:Vchi5} shows the effective potential $\hat{V}$ (normalized by the value at the critical point) plotted for various values of $\alpha$.
Here we have derived $\hat{\phi}$ by numerically solving Eq.~\eqref{eq:dphiovdhatphi}.  One can see that $\hat{\phi}_c$ (and $\hat{\phi}_*$) becomes smaller as $\alpha$ approaches $1$.
Consequently smaller $r$ is obtained, which is consistent with the results shown in Fig.~\ref{fig:ns-r_l-x_c=-5}.

\subsubsection{$\chi< -1$\,\&\,$\alpha=1$ or $\chi> -1$\,\&\,$\alpha=2/3$}
\begin{figure}[tbp]
 \begin{minipage}[t]{0.5\hsize}
   \centering
    \includegraphics[keepaspectratio,scale=0.8]{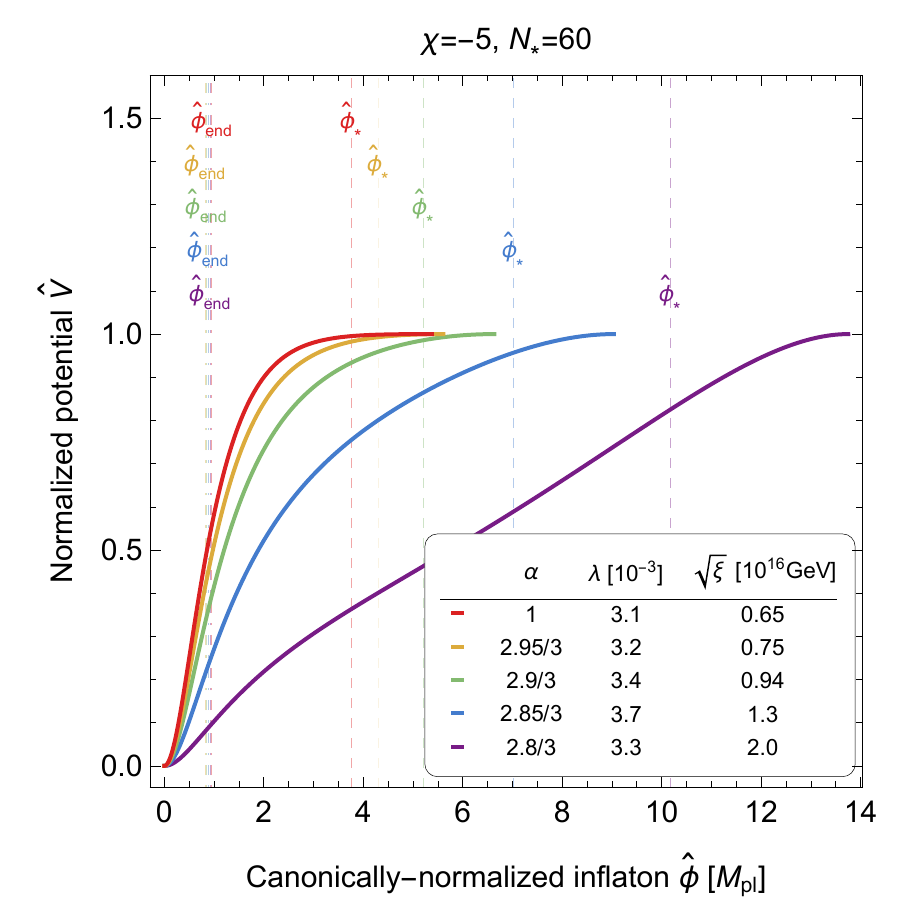}
 \end{minipage}
 \begin{minipage}[t]{0.5\hsize}
   \centering
    \includegraphics[keepaspectratio,scale=0.8]{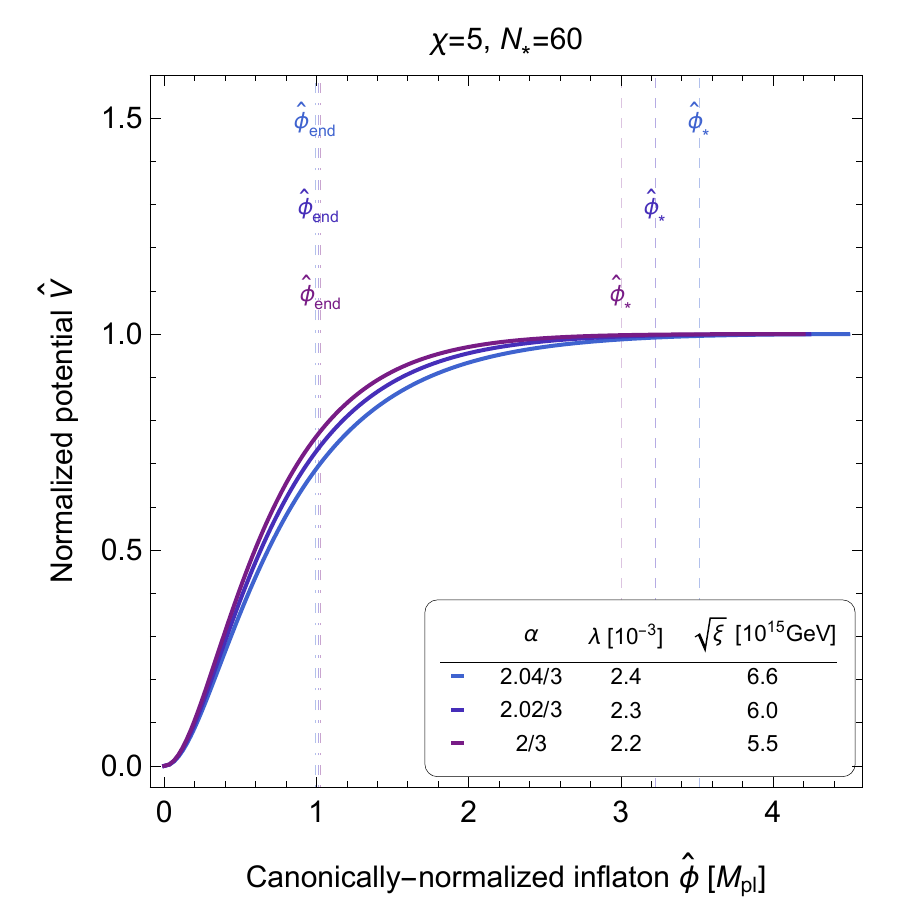}
 \end{minipage}
    \caption{Normalized potential $\hat{V}$ as a function of the canonically normalized inflaton field $\hat{\phi}$.
Only the subcritical regime is shown, and the field values of $\hat{\phi}$ at the end of inflation and at 60 $e$-folds are indicated as $\hat{\phi}_{\rm end}$ and $\hat{\phi}_{*}$, respectively. 
Color codes are the same as in Fig.~\ref{fig:ns-r_l-x_c=-5}, and $\lambda$ and $\sqrt{\xi}$ are chosen to give the best-fit value of the observed $n_s$.}
    \label{fig:Vchi5}
\end{figure}

As seen in Table \ref{table:PsiI}, $\Psi$ is given in more complicated expressions compared to the previous case; meanwhile, both the $\alpha=1$ and $2/3$ cases give a similar $\Psi$ expression.
This expression is valid when $|(1+\chi)^2\phi^2/(-2\Phi_0)|\gg 1$ is satisfied. 
We have found that the approximated $\Psi$ is valid for $\chi\lesssim -10$ and $\chi\gtrsim2$ when $\alpha=1$ and $2/3$, respectively.
Besides this, recall that $k$ has a lower bound, given in Eq.~\eqref{eq:bound_k_alpha1}. Aside from the exponential factor, $\Psi$ is given as a function of $2\alpha^2|\beta|/k$, and it has the same upper bound:
\begin{align}
  2\alpha^2|\beta|/k <1\,.
  \label{eq:validregionforalpha=1}
\end{align}
Therefore, the predicted values of $n_s$ and $r$ for $\alpha=1$ and
$2/3$ are expected to be similar to each other.
\footnote{The model with $\alpha=1$ is the same model studied in Ref.~\cite{Ishiwata:2018dxg}. The predictions for $(n_s,\,r)$ are different from those in the literature, because now we consider $\chi$ away from $-1$. } In fact, Fig.~\ref{fig:ns-r_l-x_c=-5} shows the expected results.
We have found that the lowest value of $k$ determines the maximum values for $n_s$ and $r$, and that the results resemble each other.
One can see this fact from the plot of the effective potential shown in Fig.\,\ref{fig:Vchi5}.
We will come back to $\alpha=1$ and $2/3$ cases with various values of
$\chi$ in Sec.\,\ref{sec:caseIII}.

In a nutshell, for $|\chi|\gg 1$, the potential becomes flatter as $\alpha$ approaches unity or 2/3.
This allows $\hat{\phi}_*$ to have smaller values, and consequently, we obtain smaller $r$ values.

\begin{figure}[tbp]
  \begin{minipage}[t]{0.5\hsize}
    \centering
    \includegraphics[keepaspectratio,scale=0.8]{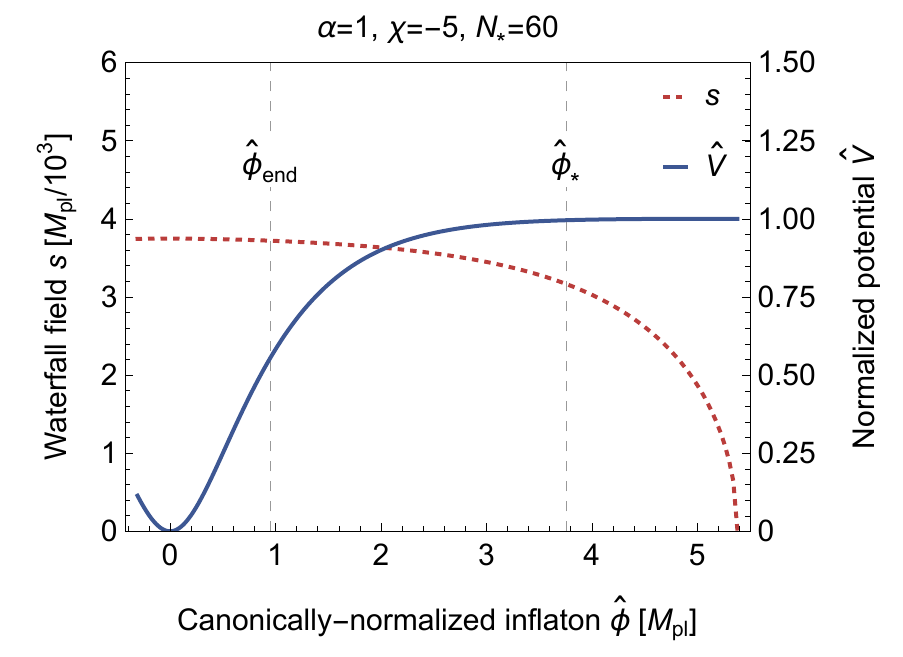}
  \end{minipage}
  \begin{minipage}[t]{0.5\hsize}
    \centering
    \includegraphics[keepaspectratio,scale=0.8]{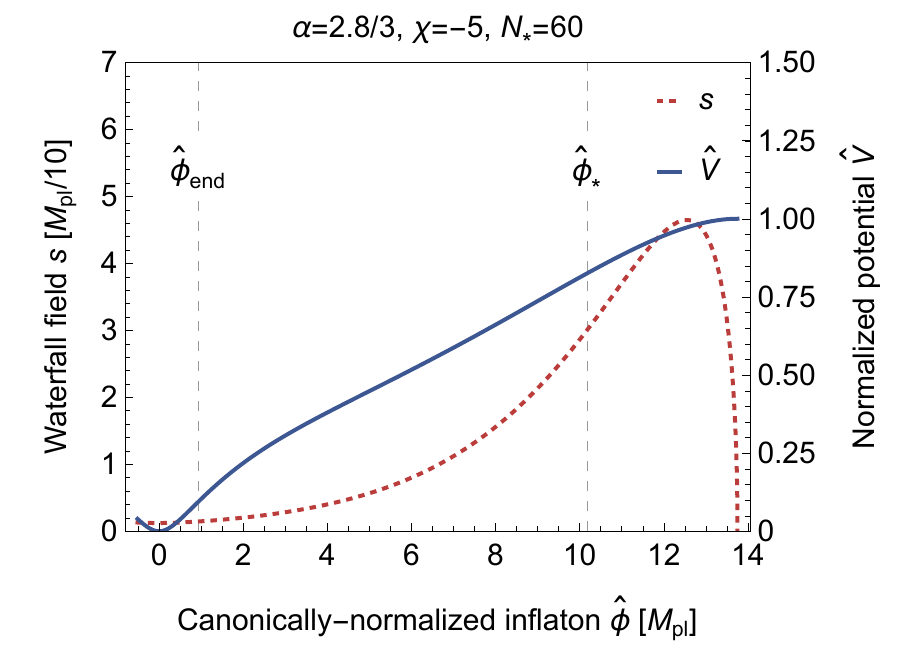}
  \end{minipage}
  \begin{minipage}[t]{0.5\hsize}
  \centering
  \includegraphics[keepaspectratio,scale=0.8]{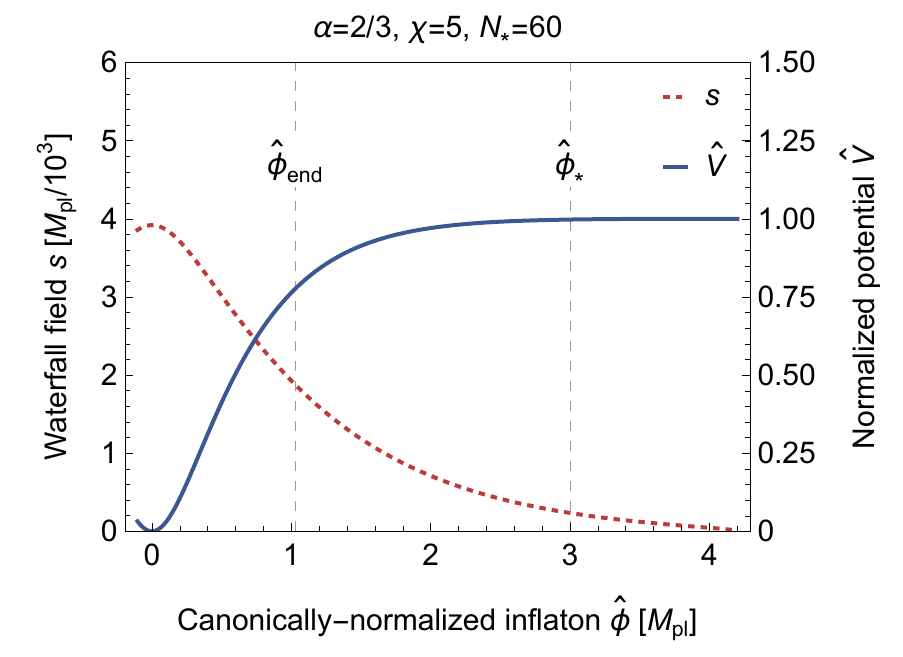}
  \end{minipage}
  \begin{minipage}[t]{0.5\hsize}
  \centering
  \includegraphics[keepaspectratio,scale=0.8]{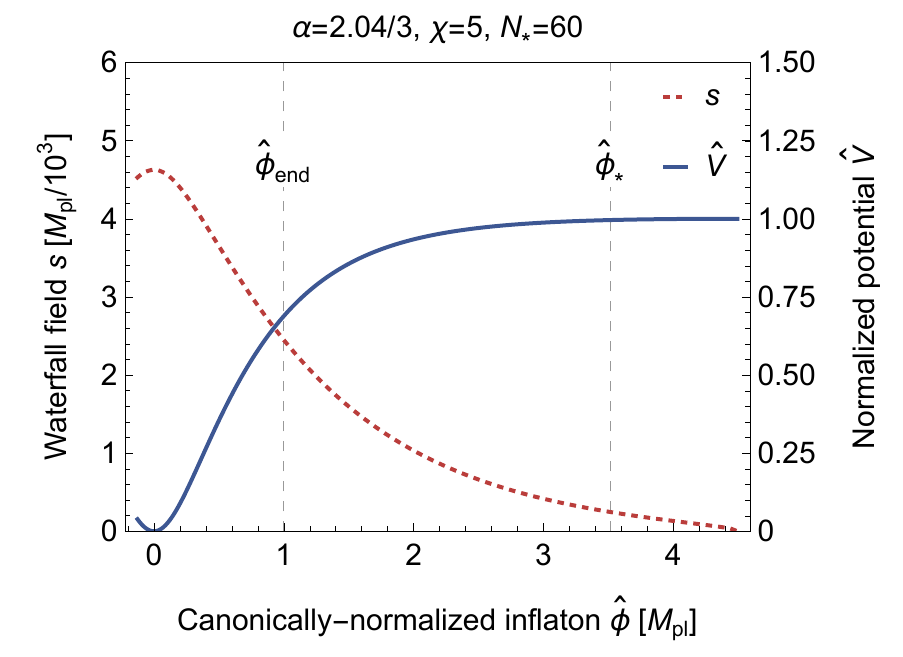}
\end{minipage}
\caption{Trajectory of the waterfall field (red dotted curve) as a function of the canonically normalized inflaton field $\hat{\phi}$.
As a reference, the normalized potential $\hat{V}$ (blue solid curve) is also plotted.
The field values of $\hat{\phi}$ at the end of inflation and at 60 $e$-folds are indicated as $\hat{\phi}_{\rm end}$ and $\hat{\phi}_{*}$, respectively.
In the top (bottom) panels, $\chi$ is taken to $-5$ ($5$). 
The other parameters are $\alpha=1$, $\lambda = 3.1\times 10^{-3}$, $\sqrt{\xi}=6.5\times 10^{15}$\,GeV (top-left); $\alpha=2.8/3$, $\lambda =3.3\times 10^{-3}$, $\sqrt{\xi}=2.0\times 10^{16}$\,GeV (top-right); $\alpha=2/3$, $\lambda = 2.2\times 10^{-3}$, $\sqrt{\xi}=5.5\times 10^{15}$\,GeV
(bottom-left); and $\alpha=2.04/3$, $\lambda = 2.4\times 10^{-3}$,
$\sqrt{\xi}=6.6\times 10^{15}$\,GeV (bottom-right).
With these parameters, the median value of the observed $n_s$ is obtained, and the tensor-to-scalar ratio is predicted to be $7.9\times 10^{-4}$
(top-left), $6.4\times 10^{-2}$ (top-right), $4.2\times 10^{-4}$
(bottom-left), and $8.5\times 10^{-4}$ (bottom-right). }
\label{fig:Vsmin_chi=-5}
\end{figure}

Finally, we plot the trajectory of the waterfall field as a function of $\hat{\phi}$ in Fig.~\ref{fig:Vsmin_chi=-5}.
The potential is found to be almost flat at $60$ $e$-folds from the end of inflation except for the $\alpha=2.8/3$ and $\chi=-5$ case.
In this case, the waterfall field grows much larger than the global minimum value due to a factor
\begin{align}
  -\frac{\Phi_0}{3}(1-\Psi) \sim
  -\Bigl(\frac{2\alpha^2\beta}{k}\Bigr)^{\frac{1}{3(1-\alpha)}}
  \sqrt{\frac{6}{\alpha}}(1-\alpha)(\hat{\phi}-\hat{\phi}_c)
\end{align}
where $(2\alpha^{2}\beta/k)^{1/3(1-\alpha)}\sim \order{10^5}$ after the critical point.
Due to this enhancement, the effective potential is distorted to have convection points.  It is worth noticing that the subcritical regime of the superconformal hybrid inflation reduces to such a potential and that predicted $n_s$ and $r$ values can be consistent with the observations. In the other cases, on the other hand, there is no such enhancement after the critical point, and the waterfall field grows slowly as shown in the figure.

\subsection{$\chi\simeq -1$}
\label{sec:caseII}

\begin{figure}[tbp]
  \begin{minipage}[t]{0.5\hsize}
  \centering
  \includegraphics[keepaspectratio,scale=0.8]{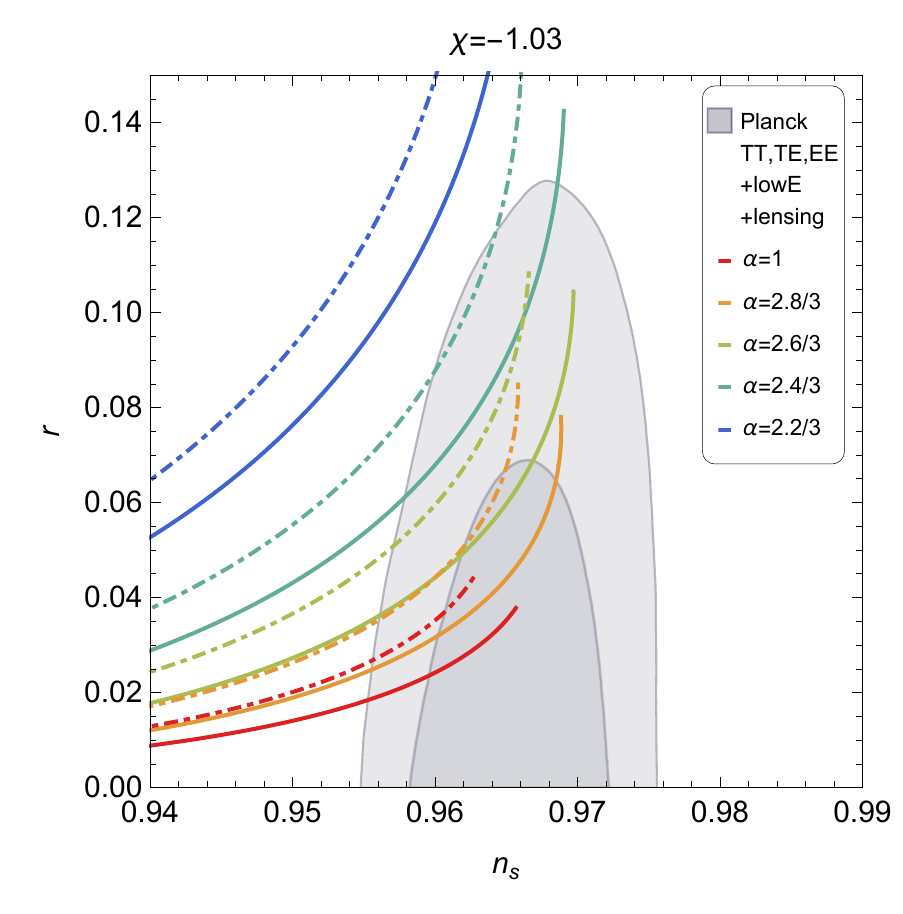}
  \end{minipage}
  \begin{minipage}[t]{0.5\hsize}
  \centering
  \includegraphics[keepaspectratio,scale=0.8]{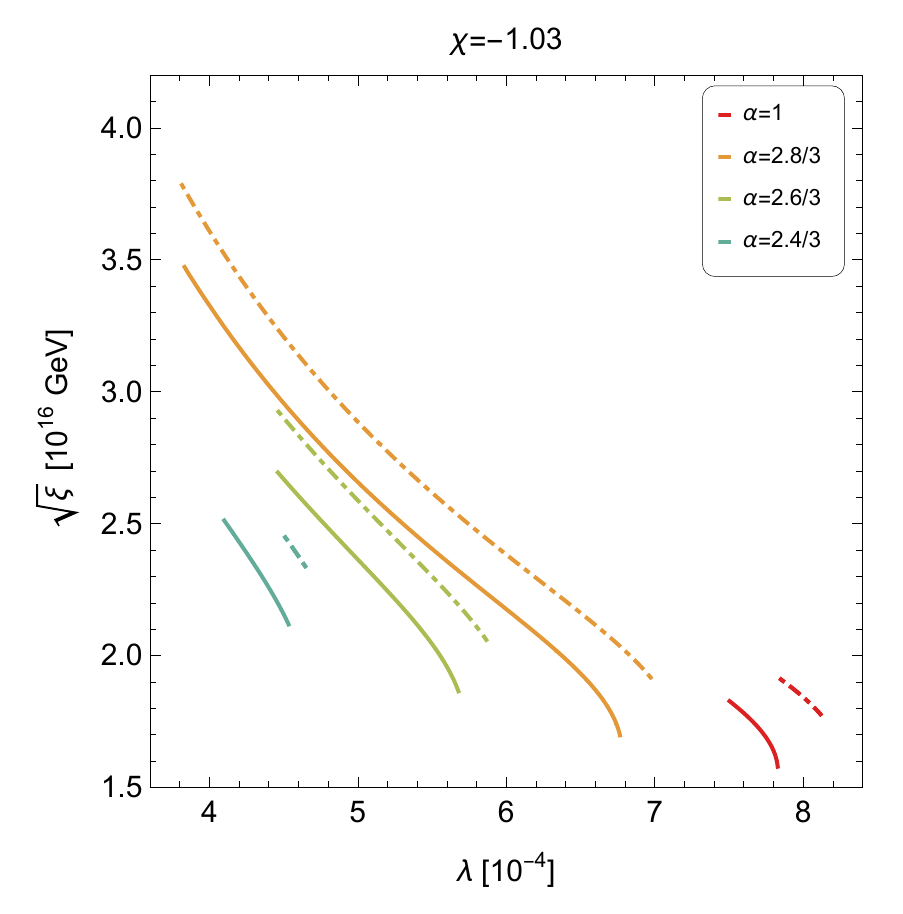}
  \end{minipage}
 \begin{minipage}[t]{0.5\hsize}
  \centering
  \includegraphics[keepaspectratio,scale=0.8]{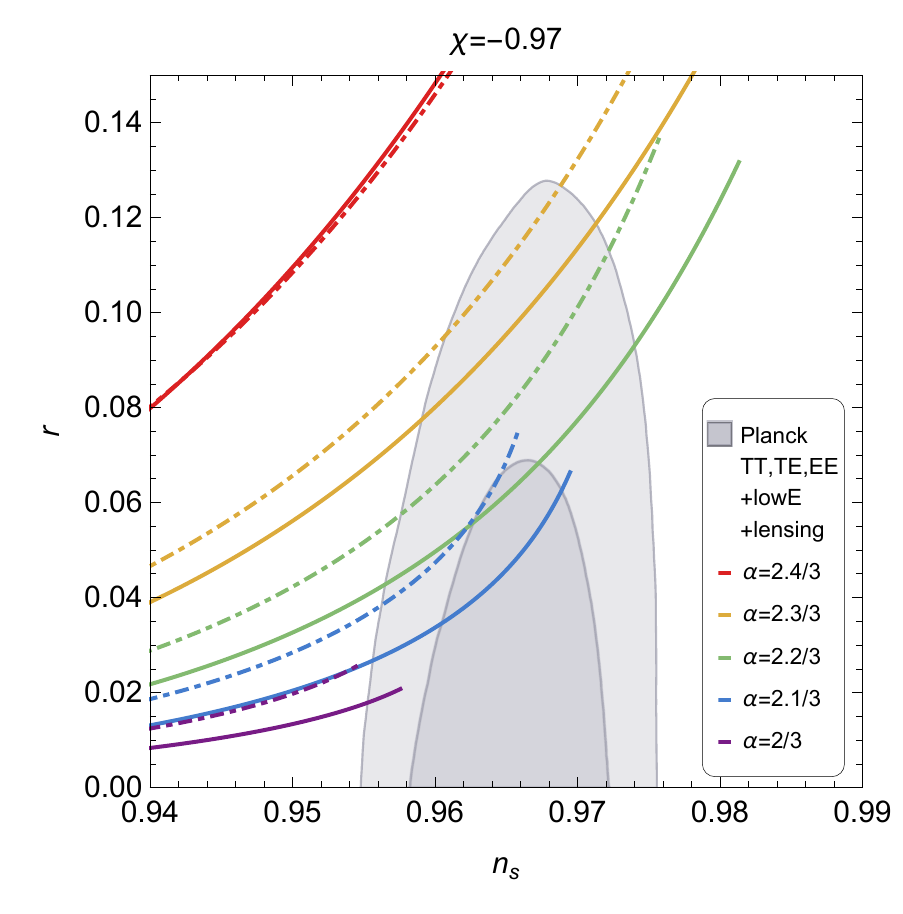}
\end{minipage}
\begin{minipage}[t]{0.5\hsize}
  \centering
  \includegraphics[keepaspectratio,scale=0.8]{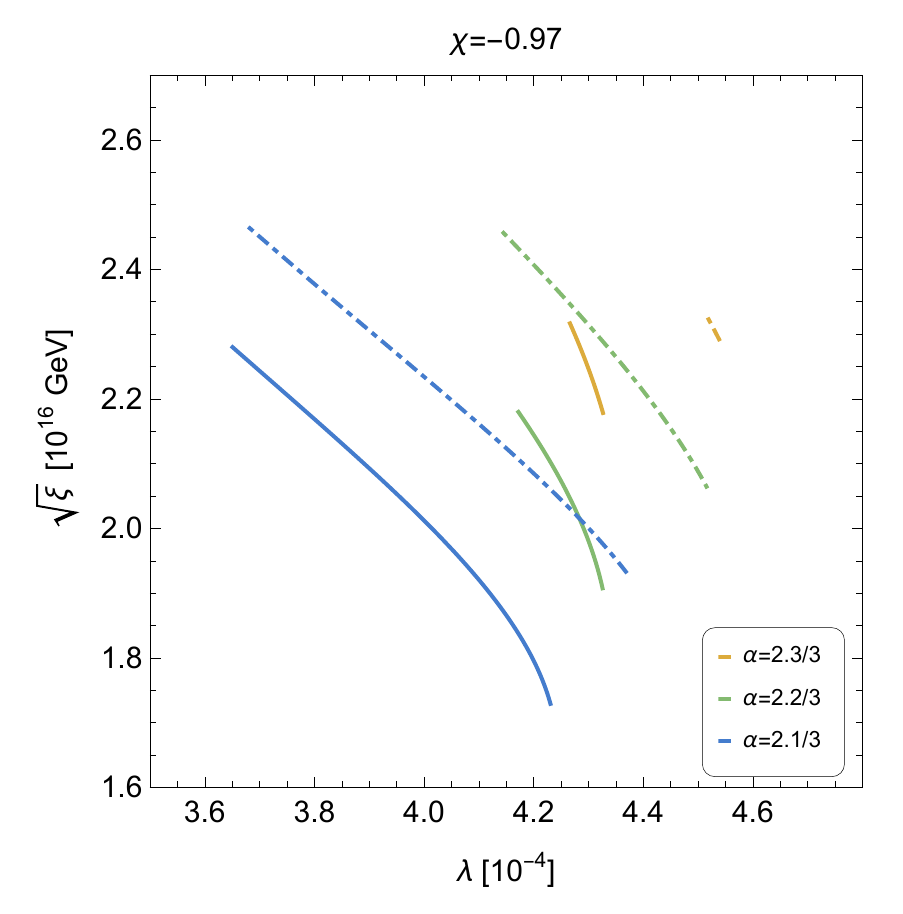}
\end{minipage}
\caption {Same as Fig.~\ref{fig:ns-r_l-x_c=-5}, but taking $\chi=-1.03$ (top) and $\chi=-0.97$ (bottom) and different values of $\alpha$ accordingly.
In the right panels, we take $N_*=60$ (solid curves) and $N_*=55$ (dot-dashed curves).}
\label{fig:ns-r_l-x_c=-1.03}
\end{figure}

\begin{figure}[tbp]
 \begin{minipage}[t]{0.5\hsize}
   \centering
    \includegraphics[keepaspectratio,scale=0.8]{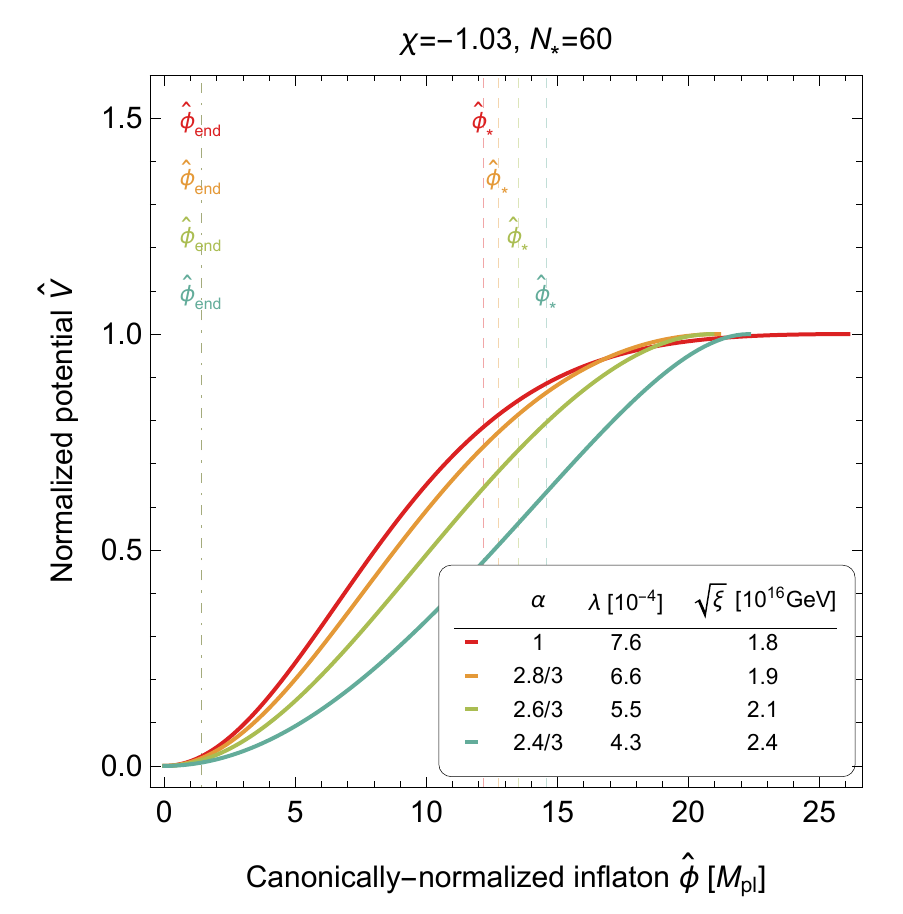}
 \end{minipage}
 \begin{minipage}[t]{0.5\hsize}
   \centering
    \includegraphics[keepaspectratio,scale=0.8]{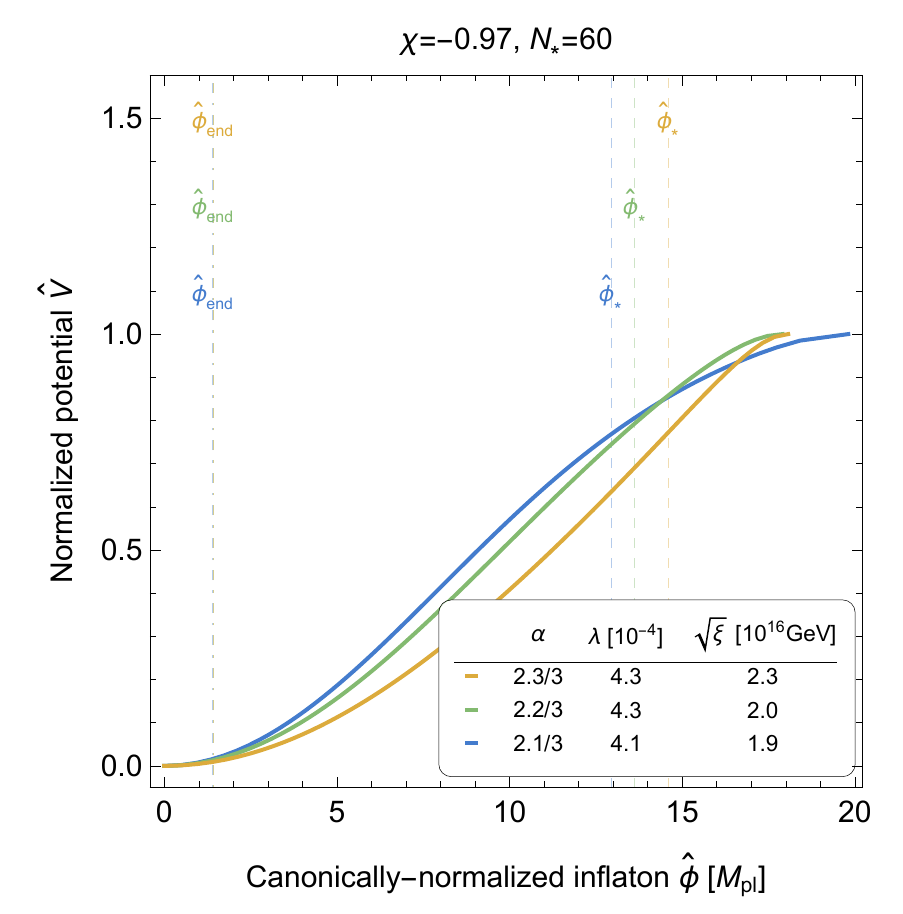}
 \end{minipage}
    \caption{Same as Fig.~\ref{fig:Vchi5} but with $\chi=-1.03$ and $-0.97$. Color codes are the same as in Fig.~\ref{fig:ns-r_l-x_c=-1.03}, and $\lambda$ and $\sqrt{\xi}$ are chosen to give the best-fit value for $n_s$.}
    \label{fig:Vchi-1}
\end{figure}

\textit{Summary of the predictions.} --- Figure \ref{fig:ns-r_l-x_c=-1.03} is the same as Fig.~\ref{fig:ns-r_l-x_c=-5}, but taking $\chi=-1.03$ (top) and $\chi=-0.97$ (bottom) and various values of $\alpha$.
In the right panels, the results are given for $N_*=60$ and $55$ (since the allowed region for $N_*=50$ is very limited).  As in the case of $|\chi|\gtrsim 5$, $n_s$ and $r$ become larger for smaller $k$. 
It is found that $10^{-2}\lesssim r\lesssim 0.1$ for both cases in the region consistent with the Planck data.
While the result with $\chi=-1.03$ is similar to one given in Ref.~\cite{Ishiwata:2018dxg}, $r$ becomes larger for smaller $\alpha$, and it is eventually out of the preferred region by the Planck observations.
The results of $\chi=-0.97$ look similar. 
However, $r$ becomes smaller for smaller $\alpha$. The allowed region is found to be $10^{-4}\lesssim\lambda \lesssim 10^{-3}$ and $\sqrt{\xi}\sim\order{10^{16}\,{\rm GeV}}$ for both cases.

As in the previous subsection, we derive the effective potential in terms of the canonically normalized field.
In the present case, the second term in the parenthesis on the rhs in Eq.~\eqref{eq:K_NN_appr} can be ignored due to  $\chi+1\simeq 0$ to get
\begin{align}
  K_{N\bar{N}}\simeq-\frac{3\alpha}{\Phi_0}\,.
\end{align}
Consequently, $\phi$ and  $\Phi_0$ are given by
\begin{align}
  \phi^{2}&\simeq 
  \begin{cases}
    \displaystyle
    \frac{1}{\beta}\sinh^{2}\sqrt{\frac{\beta}{\alpha}}\hat{\phi}
    &~~~ (\chi<-1)\\[3mm]
    \displaystyle
    \frac{1}{-\beta}\sin^{2}\sqrt{\frac{-\beta}{\alpha}}\hat{\phi}
    &~~~ (\chi>-1)
  \end{cases}\,,\\[2mm]
  -\frac{\Phi_{0}}{3}&\simeq
  \begin{cases}
    \displaystyle
    \cosh^{2}\sqrt{\frac{\beta}{\alpha}}\hat{\phi}
    &~~~ (\chi<-1)\\[3mm]
    \displaystyle
    \cos^{2}\sqrt{\frac{-\beta}{\alpha}}\hat{\phi}
    &~~~ (\chi>-1)
  \end{cases}\,,
\end{align}
and $\Psi$ is given in Table\,\ref{table:PsiII}.
Here we have taken the boundary condition $\hat{\phi}=0$ at
$\phi=0$.
One can see that both cases give similar expressions.
In fact, both are exactly the same around $\phi=0$, and the qualitative difference appears at large field values.
\footnote{As mentioned earlier, the $\chi=-1$ case corresponds to the model studied in Refs.~\cite{Buchmuller:2014rfa,Buchmuller:2014dda} when potential is written in terms of $\hat{\phi}$.}
We find that the approximated expression is valid in $|1+\chi|\lesssim 0.05$ and $|1+\chi|\lesssim 0.01$ for $\chi<-1$ and $\chi>-1$, respectively.
Since the valid parameter region is limited, it is better to see $\alpha$ dependence numerically by computing the potential, which is shown in Fig.~\ref{fig:Vchi-1}. 
From the figure, one can see that the potential becomes flatter as $\alpha$ approaches $1$ ($2/3$) and gives smaller values of $r$ for $\chi=-1.03$ ($-0.97$).
This behavior is similar to that seen in Sec.~\ref{sec:caseI}, and it can be understood qualitatively as follows.
Expanding the effective potential in terms of $\hat{\phi}$, the tensor-to-scalar ratio is approximately given by
\begin{align}
  r \sim
   \frac{32}{\hat{\phi} ^2_*}
   +\frac{32 (9 \alpha -7) (1+\chi)}{9 \alpha }\,.
    \label{eq:r_chisim-1}
\end{align}
Due to the second term, $r$ becomes smaller (larger) as $\alpha$ gets larger when $\chi<-1$ ($\chi >-1$). 
We note that the above rough estimate cannot give the precise value for $r$; however, it is enough to understand the response to the value of $\alpha$.  In addition, we find that $\alpha=2/3$ is not allowed for $\chi=-0.97$.
This comes from the lower bound for $k$ given in Eq.~\eqref{eq:bound_k_alpha1}.
We will discuss this in the next subsection in detail by comparing the results with $\alpha=1$.
Except for $\alpha=2/3$ or $1$, we find that the endpoint of $n_s$ and $r$ comes from saturated value of $\phi_*$ under $k\to 0$.

\begin{table}[t]
 \begin{center}
  \caption{\small Approximated expression of $\Psi$ in the limit $|(1+\chi)^2\phi^2/(-2\Phi_0)|\ll 1$.
The parameter space where the approximated expression is valid is given in the right column.
The effective potential can be obtained from $\Psi$ by using Eq.~\eqref{eq:inflaton_potential}.}
  \begin{tabular}{ccc}
   \hline \hline
   Cases   &  Approximated expression for $\Psi$
   & Valid parameters\\
   \hline
   $\chi< -1$
   & $
   \displaystyle
    \frac{k}{2\alpha^{2}\beta}\cosh^{2(2-3\alpha)}\sqrt{\frac{\beta}{\alpha}}\hat{\phi}
    \times\sinh^{2}\sqrt{\frac{\beta}{\alpha}}\hat{\phi}$
    & $|1+\chi|\lesssim 0.05$
    \vspace{3mm}
    \\
   $\chi> -1$
   & $
   \displaystyle
    \frac{k}{2\alpha^{2}|\beta|}\cos^{2(2-3\alpha)}\sqrt{\frac{|\beta|}{\alpha}}\hat{\phi}
    \times\sin^{2}\sqrt{\frac{|\beta|}{\alpha}}\hat{\phi}$
   & $|1+\chi|\lesssim 0.01$
   \vspace{1mm}
   \\
    \hline \hline
  \end{tabular}
  \label{table:PsiII}
 \end{center}
\end{table}

\begin{figure}[tb]
  \begin{minipage}[t]{0.5\hsize}
    \centering
    \includegraphics[keepaspectratio,scale=0.8]{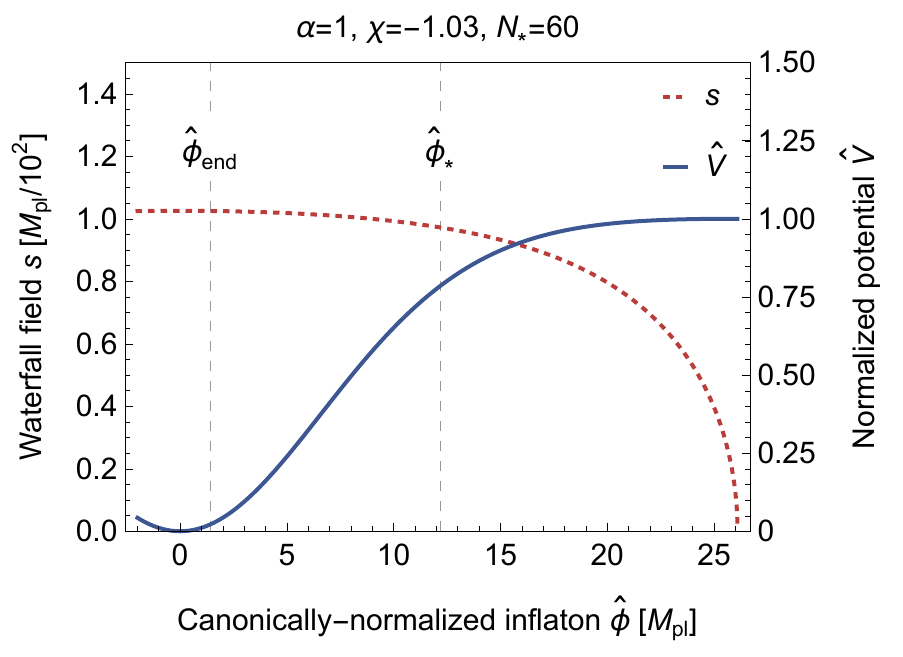}
  \end{minipage}
  \begin{minipage}[t]{0.5\hsize}
    \centering
    \includegraphics[keepaspectratio,scale=0.8]{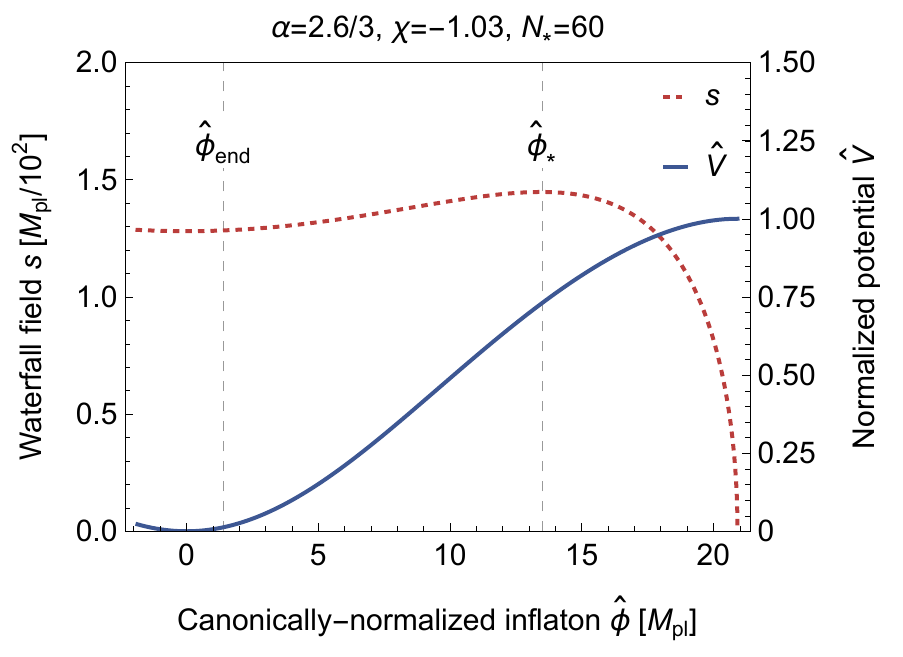}
  \end{minipage}
  \begin{minipage}[t]{0.5\hsize}
    \centering
    \includegraphics[keepaspectratio,scale=0.8]{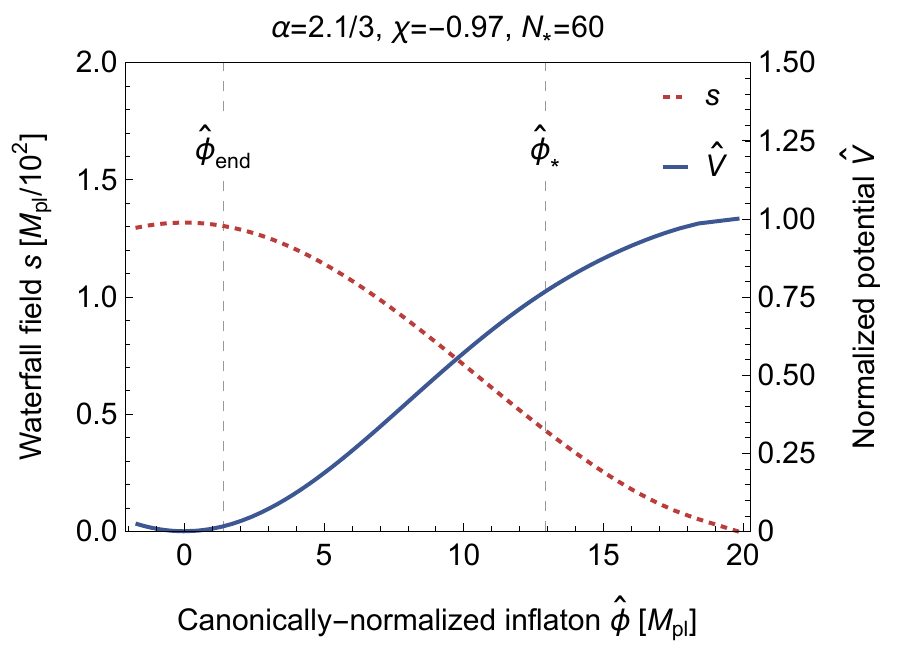}
  \end{minipage}
  \begin{minipage}[t]{0.5\hsize}
    \centering
    \includegraphics[keepaspectratio,scale=0.8]{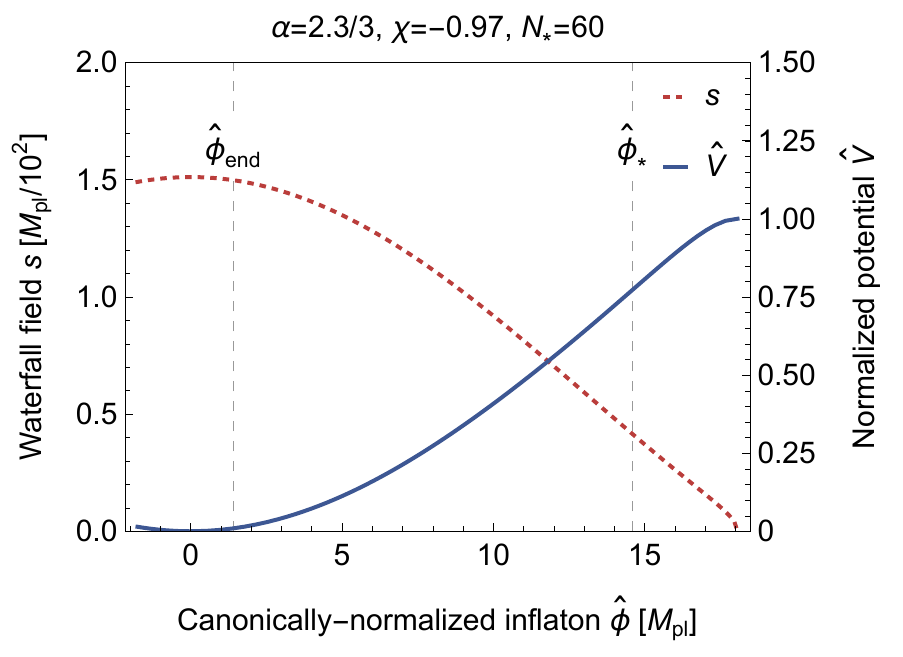}
  \end{minipage}
\caption{Same as Fig.~\ref{fig:Vsmin_chi=-5}, but taking $\chi=-1.03$ (top) and $\chi=-0.97$ (bottom).
The other parameters in each panel are $\alpha=1$, $\lambda=7.6\times 10^{-4}$, $\sqrt{\xi}=1.8\times 10^{16}$\,GeV (top-left); $\alpha=2.6/3$, $\lambda=5.5\times 10^{-4}$, $\sqrt{\xi}=2.1\times 10^{16}$\,GeV (top-right); $\alpha=2.1/3$, $\lambda=4.1\times 10^{-4}$, $\sqrt{\xi}=1.9\times 10^{16}$ GeV (bottom-left); $\alpha=2.3/3$, $\lambda=4.3\times 10^{-4}$, $\sqrt{\xi}=2.3\times 10^{16}$ GeV (bottom-right). The tensor-to-scalar ratio is found to be $3.5\times 10^{-2}$ (top-left), $5.9\times 10^{-2}$ (top-right), $4.5\times 10^{-2}$ (bottom-left), and $r=9.5\times 10^{-2}$ (bottom-right).}
\label{fig:Vsmin_chi=-1.03}
\end{figure}

For comparison with the $|\chi|\gtrsim 5$ case, Fig.~\ref{fig:Vsmin_chi=-1.03} shows the trajectory of the waterfall field.
Although the potential is not so flat, a large field value keeps the inflaton slow-roll.
For $\chi<-1$, the waterfall field grows to its global minimum value just after entering the subcritical regime, but it does not overshoot much greater than the global minimum.

\subsection{Specific cases}
\label{sec:caseIII}

\begin{figure}[tbp]
  \begin{minipage}[t]{0.5\hsize}
    \centering
    \includegraphics[keepaspectratio,scale=0.8]{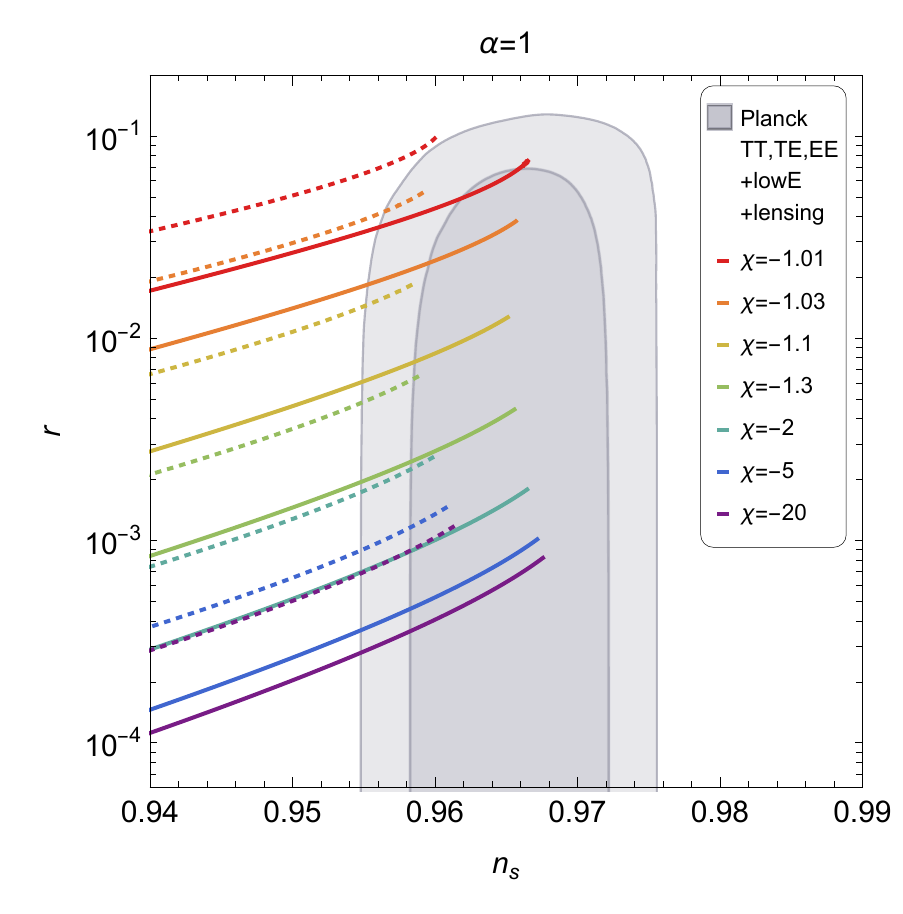}
  \end{minipage}
  \begin{minipage}[t]{0.5\hsize}
    \centering
    \includegraphics[keepaspectratio,scale=0.8]{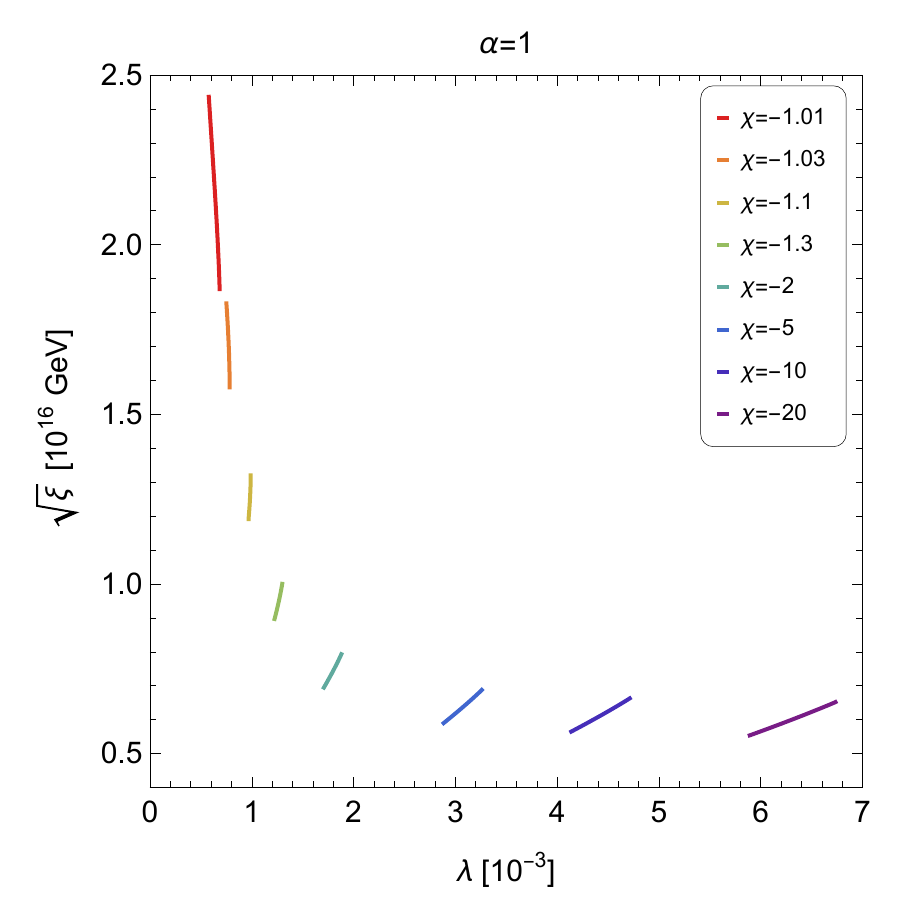}
  \end{minipage}
  \begin{minipage}[t]{0.5\hsize}
    \centering
    \includegraphics[keepaspectratio,scale=0.8]{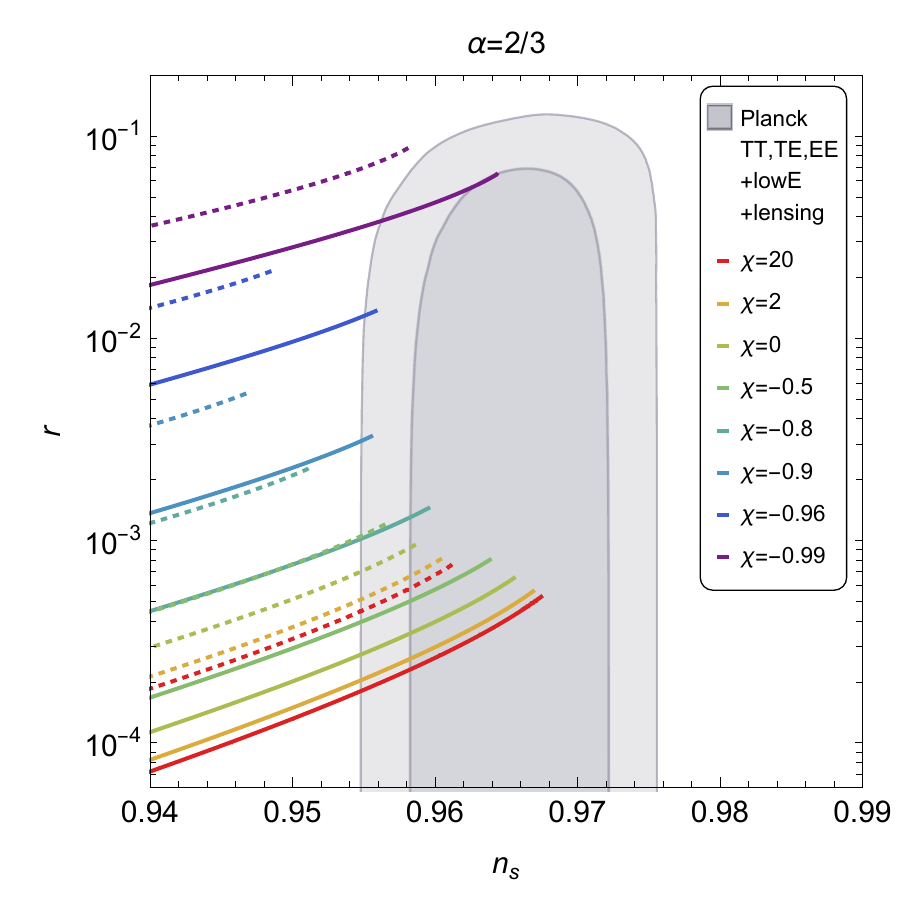}
  \end{minipage}
  \begin{minipage}[t]{0.5\hsize}
    \centering
    \includegraphics[keepaspectratio,scale=0.8]{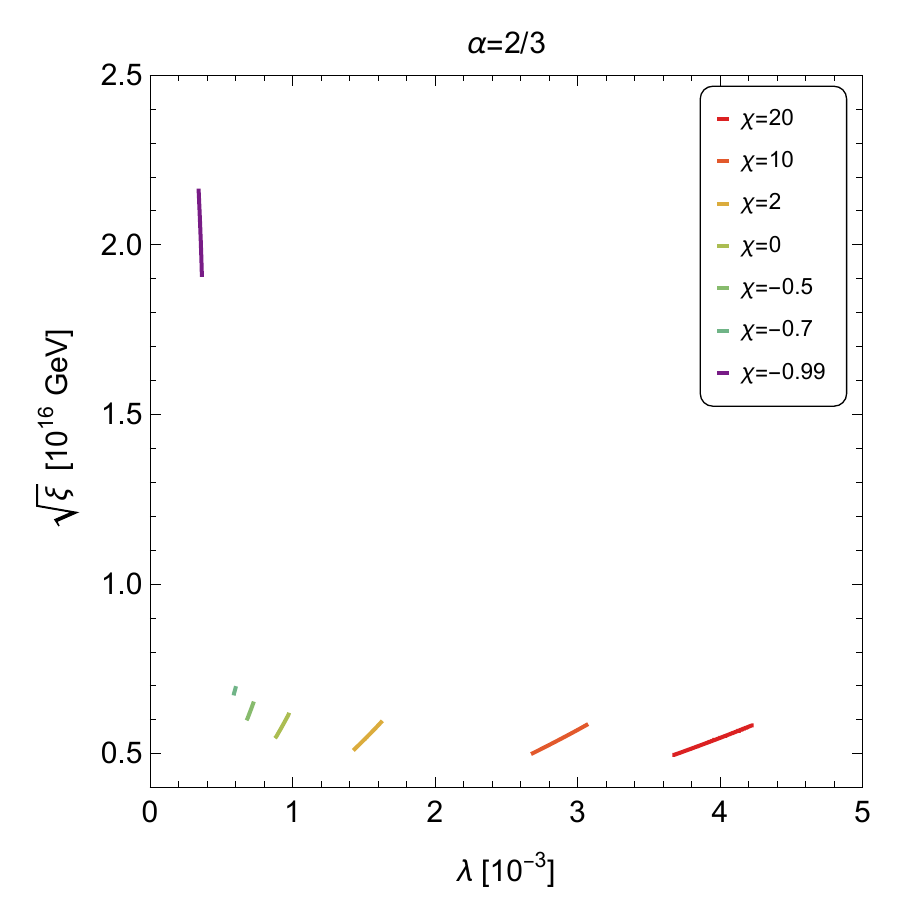}
  \end{minipage}
  \caption{\textit{Left:} Predicted values of $n_s$ and $r$ for $N_*=60$ (solid curves) and $50$ (dotted curves) and various values of $\chi$.  $\alpha$ is taken to be $1$ (top) and $2/3$ (bottom).
\textit{Right:} Allowed parameters for $N_*=60$. Here, $\alpha=1$ (top) and $2/3$ (bottom), and various values of $\chi$ are taken.   } 
\label{fig:nsr_lx_a=1_2ov3}
\end{figure}

\begin{figure}[tbp]
 \begin{minipage}[t]{0.5\hsize}
   \centering
    \includegraphics[keepaspectratio,scale=0.8]{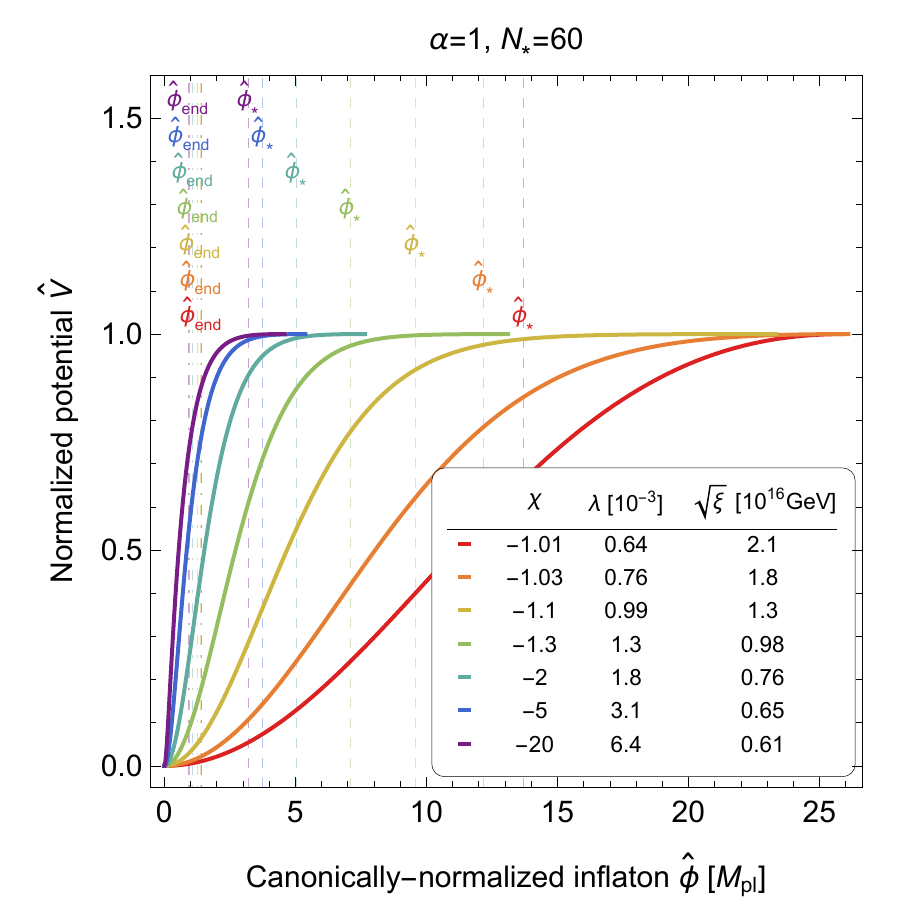}
 \end{minipage}
 \begin{minipage}[t]{0.5\hsize}
   \centering
    \includegraphics[keepaspectratio,scale=0.8]{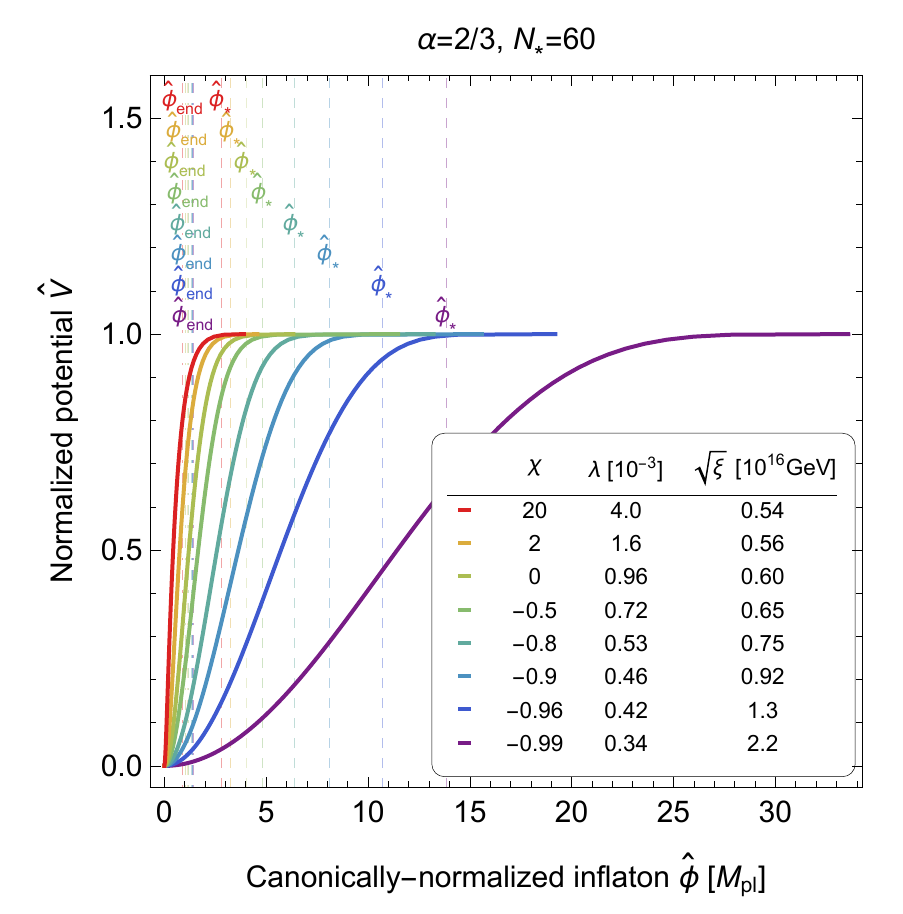}
 \end{minipage}
    \caption{Normalized potential as a function of the canonically normalized inflaton field.
Color codes are the same as in Fig.~\ref{fig:nsr_lx_a=1_2ov3}, and $\lambda$ and $\sqrt{\xi}$ are chosen to give the best-fit value for $n_s$.}
    \label{fig:Valpha1}
\end{figure}

Finally, we focus on the specific cases where $\alpha=1$, $\alpha=2/3$, and $\chi=0$. Such values of the parameters are motivated by superstring theory or superconformal symmetry.
Therefore, it is worth analyzing these cases in detail, although part of the results shown in this subsection are already presented in the previous subsections.
As is found in the previous subsections, the results for $\alpha=1$ and $2/3$ share some behaviors.  Thus we categorize the contents as $\alpha=1$ or $2/3$, and $\chi=0$.

\subsubsection{$\alpha=1$ or $2/3$}

\textit{Summary of the predictions.} --- Figure \ref{fig:nsr_lx_a=1_2ov3} shows the predicted $n_s$ and $r$ (left panels) and the allowed region (right panels) for $\alpha=1$ (top) and $\alpha=2/3$ (bottom) and various values of $\chi$.
The allowed parameters are roughly $\lambda\sim \order{10^{-3}}$ and $\sqrt{\xi}\sim \order{10^{16}\,{\rm GeV}}$ and mildly depend on $\chi$. 
For $\alpha=1$, $\chi<-1$ gives consistent results with the Planck observations for $N_*=60$, while $N_*=50$ is found to be disfavored. For $\alpha=2/3$, $n_s$ and $r$ are consistent with the observed data when $\chi \gtrsim 0$ or $\chi \simeq -1$.
In the latter case, $r\sim 0.1$, and $N_*=60$ is preferred.
As in $\alpha=1$ case, $N_*=50$ has a tension with the observed data.

It can be seen the resultant $n_s$ and $r$ for $\alpha=1$ and $2/3$ are similar when $\chi\simeq -1$ and $|\chi|\gg 1$.
The similarity at $\chi\simeq -1$ is due to the fact that the effective potential reduces to the same potential as shown in Table~\ref{table:PsiII}.
In the $|\chi|\gg 1$ case, $\Psi$ has the same form of function but with a different exponent of the exponential factor (see Table~\ref{table:PsiI}) and the valid domain of the parameter.
That is why the behavior is quite similar. However, they give quantitatively different predictions for $n_s$ and $r$, which comes from the different exponent in $\Psi$. The quantitative difference is prominent in the other value of $\chi$: {\it i.e.}, $\chi\lesssim -1$ and $\chi \gtrsim -1$.

Fig.~\ref{fig:Valpha1} shows $\chi$ dependence on
the potential.
It is clearly seen that the potential gets flatter
for larger value of $|\chi|$ for both $\alpha=1$ and $2/3$, and
consequently $r$ is suppressed.

\begin{figure}[tb]
  \begin{minipage}[t]{0.5\hsize}
    \centering
    \includegraphics[keepaspectratio,scale=0.8]{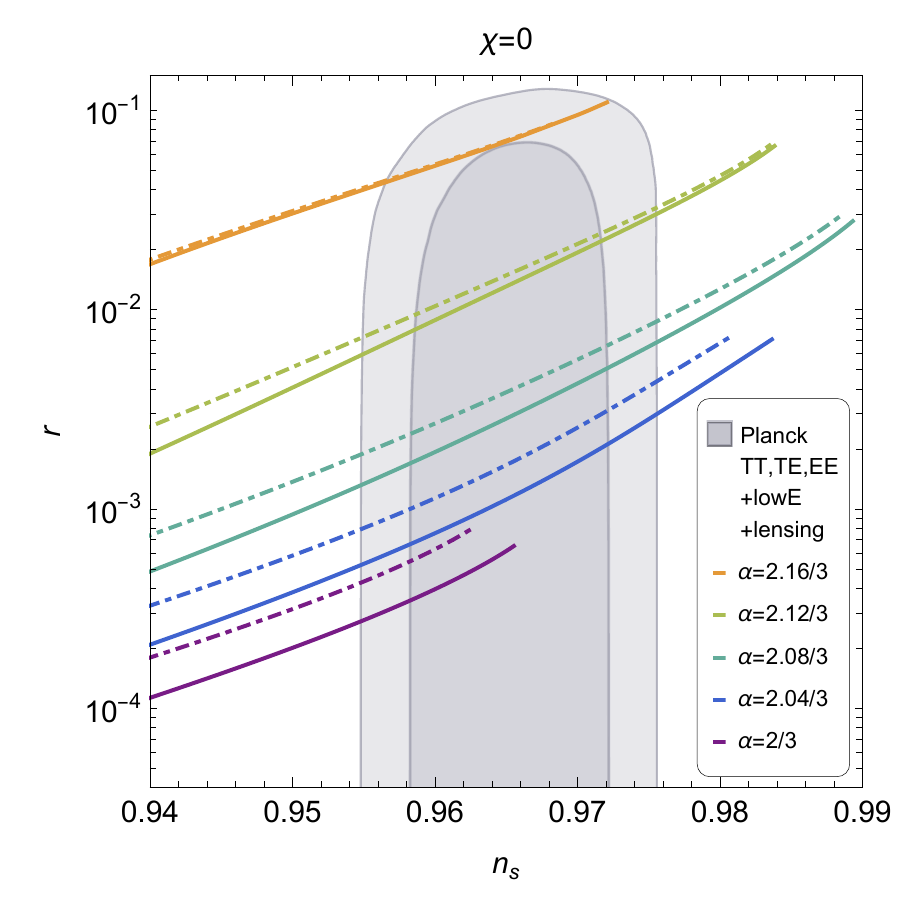}
  \end{minipage}
  \begin{minipage}[t]{0.5\hsize}
    \centering
    \includegraphics[keepaspectratio,scale=0.8]{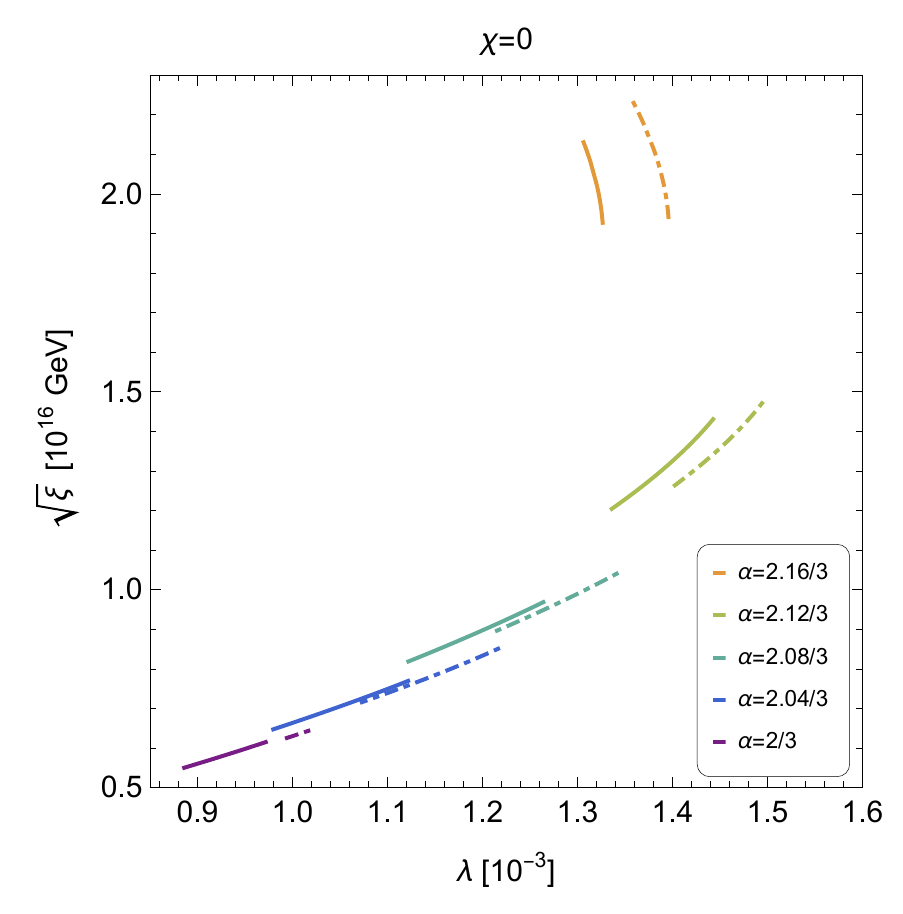}
  \end{minipage}
\caption{Same as Fig.~\ref{fig:ns-r_l-x_c=-5}, but taking $\chi=0$ and different values of $\alpha$ accordingly.
  In the right panels, we take $N_*$=60 (solid curves) and $N_*=55$ (dot-dashed curves).
}
\label{fig:nsr_lx_c=0}
\end{figure}

\begin{figure}[tbp]
  \centering
    \includegraphics[keepaspectratio,scale=0.8]{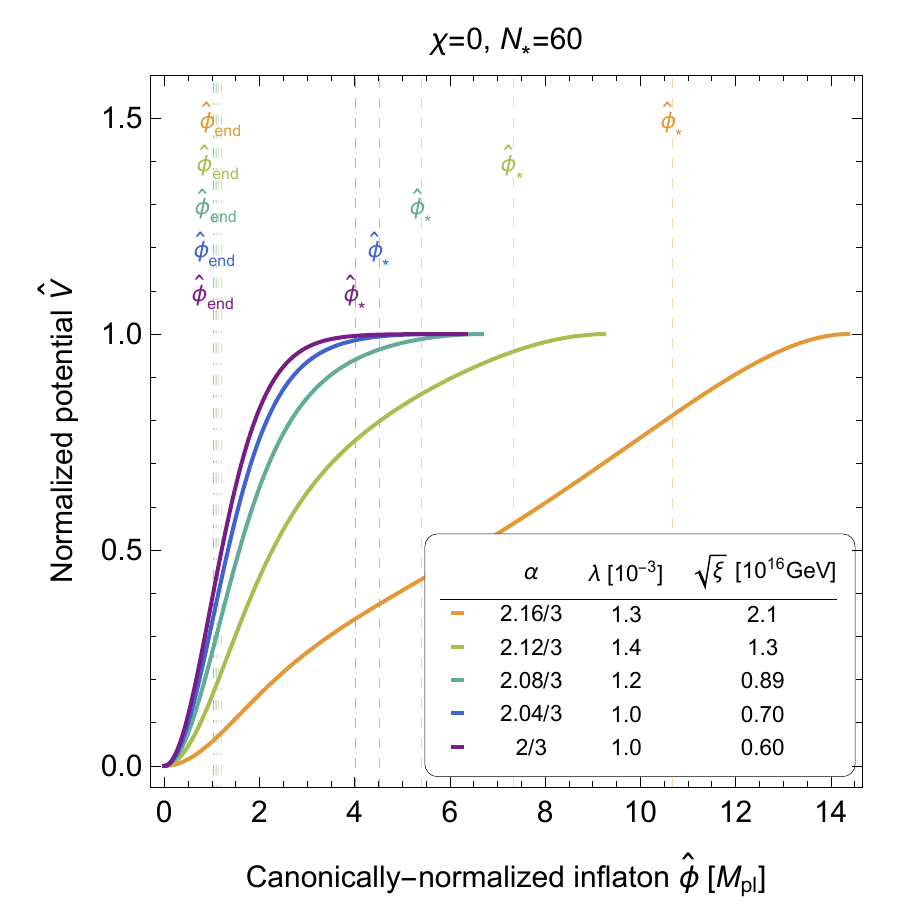}
    \caption{Same as Fig.~\ref{fig:Vchi5} but with $\chi=0$.  Color codes are the same as in Fig.~\ref{fig:nsr_lx_c=0}, and $\lambda$ and $\sqrt{\xi}$ are chosen to give the best-fit value for $n_s$.}
    \label{fig:Vchi0}
\end{figure}

\begin{figure}[tb]
  \begin{minipage}[t]{0.5\hsize}
  \centering
  \includegraphics[keepaspectratio,scale=0.8]{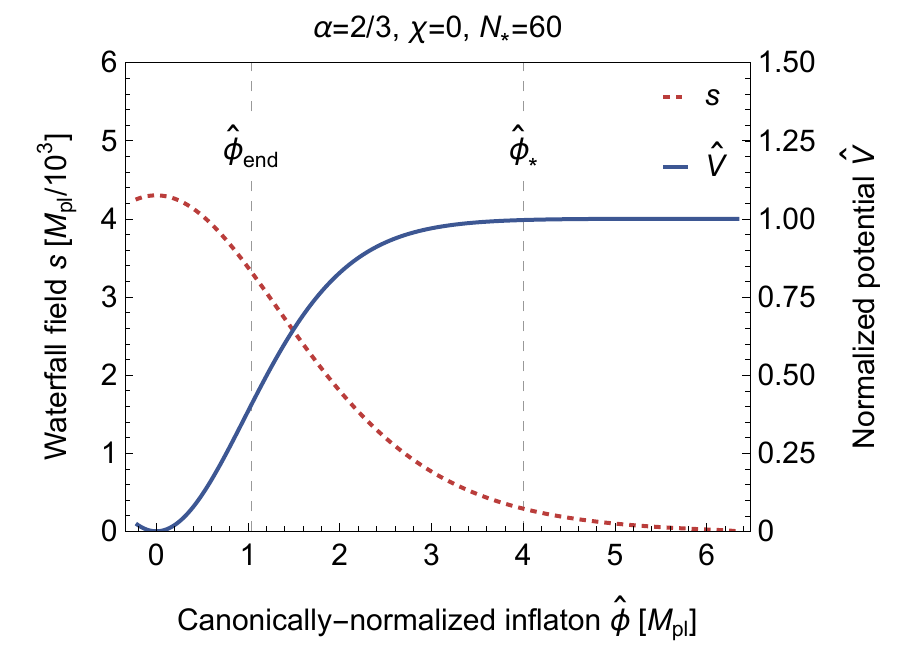}
  \end{minipage}
  \begin{minipage}[t]{0.5\hsize}
  \centering
  \includegraphics[keepaspectratio,scale=0.8]{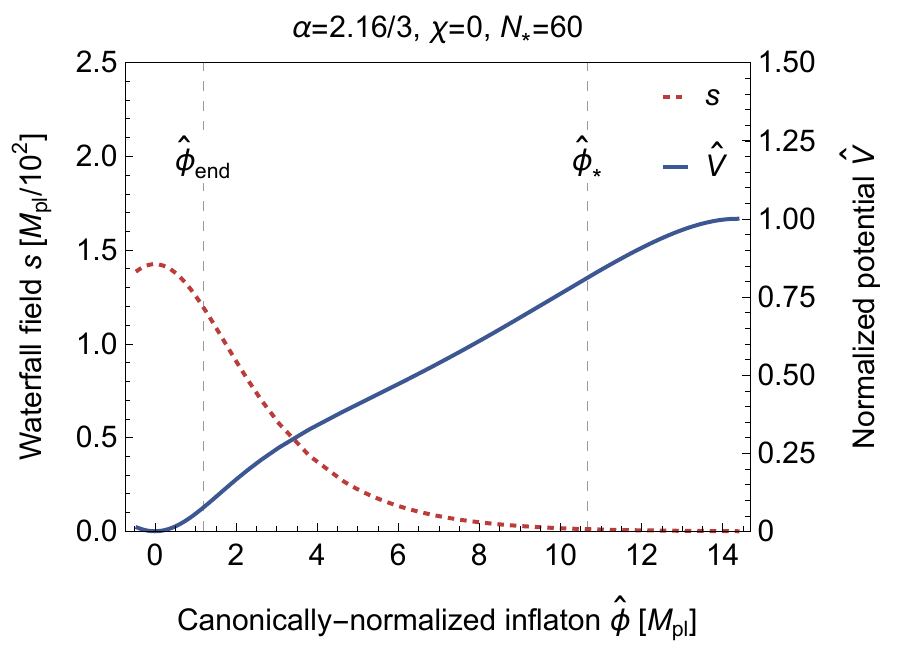}
\end{minipage}
  \caption{Same as Fig.~\ref{fig:Vsmin_chi=-5}, but taking $\chi=0$ and $\alpha=2/3$ (left) and $2.16/3$ (right).
In the left panel, $\lambda = 9.6\times 10^{-4}$ and $\sqrt{\xi}=6.0\times 10^{15}$\,GeV.
In the right panel, $\lambda = 1.3\times 10^{-3}$ and $\sqrt{\xi}=2.1\times 10^{16}$\,GeV.  The tensor-to-scalar ratio is $6.1\times 10^{-4}$ (left) and $7.0\times 10^{-2}$ (right).}
  \label{fig:Vsmin_chi=0}
\end{figure}

\subsubsection{$\chi=0$}

\textit{Summary of the predictions.} --- Fig.~\ref{fig:nsr_lx_c=0} shows the $n_s$, $r$, and the allowed parameter space for $\chi=0$.
Roughly speaking, the results are found to be similar to those in the $\chi\gtrsim 5$ case (see Fig.\,\ref{fig:ns-r_l-x_c=-5}). 
It is found that $\alpha\lesssim 2.16/3$ gives a consistent result with the observations for both $N_*=60$ and $50$.
A notable difference is that $r$ can be as large as $\order{10^{-2}}$.

In the $\chi=0$ case, it is easy to derive the potential analytically, since the K\"{a}hler metric [Eq.~\eqref{eq:K_NN_appr}] takes a simple form without any approximation:
 \begin{align}
   K_{N\bar{N}}=\frac{36\alpha}{(6-\phi^{2})^{2}}\,.
 \end{align}
 As a result, Eq.~\eqref{eq:dphiovdhatphi} can be solved analytically to give
 \begin{align}
   \phi^{2}&=
   6\tanh^{2}\frac{\hat{\phi}}{\sqrt{6\alpha}}\,,\\
   -\frac{\Phi_{0}}{3}&= 
   \cosh^{-2}\frac{\hat{\phi}}{\sqrt{6\alpha}}\,,\\
   \Psi&=
   \frac{3k}{\alpha^{2}}\tanh^{2}\frac{\hat{\phi}}{\sqrt{6\alpha}}
   \times\cosh^{2(3\alpha-2)}\frac{\hat{\phi}}{\sqrt{6\alpha}}\,.
 \end{align}
 Here, we have taken the boundary condition $\hat{\phi}=0$ at $\phi=0$. Although the effective potential is given exactly, it behaves nontrivially as a function of $\alpha$.
When $\alpha$ is away from $2/3$, the discussion for $\chi\gg -1$ case in Sec.~\ref{sec:caseI} can be applied.
Namely, when $\alpha$ gets smaller, $r$ becomes smaller. If $\alpha$ goes much closer to $2/3$, on the other hand, Eq.~\eqref{eq:condforapp} is no longer satisfied.
Instead, we can derive the effective potential for $\alpha=2/3$ as
 \begin{align}
   V=\frac{27 g^2\xi^2 k}{4} \tanh^{2}\frac{\hat{\phi}}{2}\times
   \Bigl(1-\frac{27k}{8} \tanh^{2}\frac{\hat{\phi}}{2}\Bigr)\,.
 \end{align}
The largest values of $n_s$ and $r$ are obtained from the lower bound for $k$. The resultant $n_s$ and $r$ that are consistent with the observed data are found for $N_*=60$ and $50$.
In order to see $\alpha$ dependence on $n_s$ and $r$, it is more intuitive to plot the effective potential, which is shown in Fig.~\ref{fig:Vchi0}.
As $\alpha$ approaches 2/3, it can be seen that the potential becomes flatter, and smaller values of $r$ are obtained.

For completeness, Fig.~\ref{fig:Vsmin_chi=0} shows the same plot as Fig.~\ref{fig:Vsmin_chi=-5} but takes $\chi=0$ and $\alpha=1$, $2.16/3$.
It can be seen that the shape of the potential changes nontrivially depending on $\alpha$.
That is why the prediction for $r$ changes by orders of magnitude for an $\alpha$ that even slightly deviates from $2/3$.

\section{Conclusions and discussion}
\label{sec:conclusion}

We have studied the subcritical regime of $D$-term hybrid inflation in the generalized framework of a superconformal model.
The model is characterized by the superconformal K\"{a}hler potential and
superconformal superpotential.
The former contains a parameter $\alpha$ and an explicit superconformal breaking term that is turned on by nonzero $\chi$.
The latter is given by the Yukawa interaction of the inflaton and waterfall fields with a coupling $\lambda$.
In addition, we introduce the Fayet-Iliopoulos term $\xi$ that appears after gauge fixing of the superconformal symmetry.
In this framework, we focus on the parameter space $\lambda \ll 1$, which is supported by an approximate shift symmetry of the inflaton field, and $\alpha\le 1$ ($\alpha\ge 2/3$) for $\chi<-1$ ($\chi>-1$) to give a single critical point.

In the parameter space, it has been found that inflation continues in the subcritical regime of the inflaton field, and that the inflaton potential in the subcritical regime changes drastically depending on $\alpha$ and $\chi$.
The latest Planck data prefer $2/3\le \alpha \le 1$ for $\chi\simeq -1$ and $\alpha\simeq 1$ ($2/3$) for $\chi \lesssim -1 $ ($\chi\gtrsim-1$) for 60 $e$-folds, while the preferred parameter space is limited for $50$ $e$-folds as $2.5/3\lesssim\alpha\le1$ ($2/3\le\alpha\lesssim2.2/3$) for $\chi<-1$ ($\chi>-1$).
The tensor-to-scalar ratio $r$ turns out to be $r>\order{10^{-2}}$ for $\chi\simeq -1$, $r>\order{10^{-3}}$ for $\chi\lesssim -1$, and $r\sim \order{10^{-4}}$ for $\chi\gtrsim -1$ for $60$ $e$-folds.
Roughly speaking, $r$ tends to be suppressed when $\alpha$ approaches $1$ for $\chi<-1$, or $2/3$ for $\chi>-1$ or $|\chi|\gg 1$.

The other parameters, $\lambda$ and $\xi$, on the other hand, are $\order{10^{-4}}<\lambda<\order{10^{-2}}$ and $\sqrt{\xi}\sim \order{10^{16}\,{\rm GeV}}$.
This result indicates that the FI term is determined to be around the GUT scale, which might be a clue for further phenomenological study of the GUT, neutrino sector, and inflation~\cite{Domcke:2014zqa,Domcke:2017xvu,Domcke:2017rzu,Gunji:2019wtk}.

Besides this, the cases of integer $3\alpha$ (namely, $2$ or $3$ in the present model) and $\chi=0$ are motivated by the compactification of the extra dimensions in superstring theory and the superconformal symmetry, respectively. 
We have found the allowed parameter spaces for such cases.
This may bring another clue to investigating the relation between the symmetry of the compactified space and extension of the minimal supersymmetric standard model that accommodates the inflaton sector.

As mentioned in the Introduction, non-Abelian discrete symmetry can be one of such symmetries.
Recently, the modular symmetry has drawn a lot of attention~\cite{Feruglio:2017spp} giving a nice fit with the experimental results of neutrino oscillations---for example, under the modular $S_3$~\cite{Kobayashi:2018vbk}, $A_4$~\cite{Feruglio:2017spp,Kobayashi:2018vbk,Criado:2018thu,Kobayashi:2018scp,Novichkov:2018yse,Nomura:2019jxj,Kobayashi:2019mna,Ding:2019zxk,Kobayashi:2019xvz,Kobayashi:2019gtp}, $S_4$~\cite{Penedo:2018nmg,Novichkov:2018ovf,King:2019vhv}, and $A_5$~\cite{Novichkov:2018nkm,Ding:2019xna}.
Furthermore, the study of the modular symmetry has been applied to solve the cosmological issues. Reference \cite{Asaka:2019vev} has studied leptogenesis with the modular $A_4$ and showed that right-handed neutrinos with a mass scale of $10^{13}$~GeV can account for the observed baryon asymmetry of the Universe.
This coincides with the mass scale of right-handed sneutrinos that plays the role of inflaton and can be a source of baryon asymmetry in the superconformal framework embedded into the MSSM~\cite{Ishiwata:2018dxg,Gunji:2019wtk}. 
Since the model proposed in Ref.~\cite{Gunji:2019wtk} predicts that one of the light neutrinos is massless, the model may give completely different consequences on inflation and the leptogenesis if the modular symmetry rules the lepton sector.
Additionally, it was pointed out that higher-dimension operators that are allowed by the modular symmetry in the K\"ahler potential have the possibility to spoil the success of fitting with the neutrino oscillation data~\cite{Chen:2019ewa}.
To avoid such a danger, a large volume limit in superstring theory~\cite{Kaplunovsky:1995jw,Antoniadis:1994hg} is considered.
For instance, the model with an additional gauge singlet Higgs in the large volume limit can give a consistent result with the neutrino oscillation data~\cite{Asaka:2020tmo}.
However, whether it is compatible with the cosmological issues needs further investigation.
Our results, especially on the symmetry-enhanced points, would be another possibility to provide a phenomenologically and cosmologically acceptable scenario~\cite{work-in-progress}.

\section*{Acknowledgments}

We are grateful to Tatsuo Kobayashi and Tomo Takahashi for valuable discussions.
We also thank Takashi Shimomura for hosting the ``Miyazaki Workshop on Particle Physics and Cosmology 2020'', where this work was initiated.
This work was supported by JSPS KAKENHI Grant No. JP17K14278, No. JP17H02875, No. JP18H05542, and No. JP20H01894 (K.I.).

\newpage
\bibliography{draft}

\begin{thebibliography}{10}

\bibitem{Aghanim:2018eyx}
Planck, N.~Aghanim {\em et~al.},
\newblock Astron. Astrophys. {\bf 641}, A6 (2020), arXiv:1807.06209.

\bibitem{Akrami:2018odb}
Planck, Y.~Akrami {\em et~al.},
\newblock Astron. Astrophys. {\bf 641}, A10 (2020), arXiv:1807.06211.

\bibitem{Starobinsky:1980te}
A.~A. Starobinsky,
\newblock Phys. Lett. B {\bf 91}, 99 (1980).

\bibitem{Mukhanov:1981xt}
V.~F. Mukhanov and G.~V. Chibisov,
\newblock JETP Lett. {\bf 33}, 532 (1981).

\bibitem{Kallosh:2013yoa}
R.~Kallosh, A.~Linde, and D.~Roest,
\newblock JHEP {\bf 11}, 198 (2013), arXiv:1311.0472.

\bibitem{Buchmuller:2013zfa}
W.~Buchmuller, V.~Domcke, and K.~Kamada,
\newblock Phys. Lett. B {\bf 726}, 467 (2013), arXiv:1306.3471.

\bibitem{Buchmuller:2014rfa}
W.~Buchmuller, V.~Domcke, and K.~Schmitz,
\newblock JCAP {\bf 11}, 006 (2014), arXiv:1406.6300.

\bibitem{Buchmuller:2014dda}
W.~Buchmuller and K.~Ishiwata,
\newblock Phys. Rev. D {\bf 91}, 081302 (2015), arXiv:1412.3764.

\bibitem{Clesse:2010iz}
S.~Clesse,
\newblock Phys. Rev. D {\bf 83}, 063518 (2011), arXiv:1006.4522.

\bibitem{Kodama:2011vs}
H.~Kodama, K.~Kohri, and K.~Nakayama,
\newblock Prog. Theor. Phys. {\bf 126}, 331 (2011), arXiv:1102.5612.

\bibitem{Clesse:2012dw}
S.~Clesse and B.~Garbrecht,
\newblock Phys. Rev. D {\bf 86}, 023525 (2012), arXiv:1204.3540.

\bibitem{Freese:1990rb}
K.~Freese, J.~A. Frieman, and A.~V. Olinto,
\newblock Phys. Rev. Lett. {\bf 65}, 3233 (1990).

\bibitem{Mikura:2020qhc}
Y.~Mikura, Y.~Tada, and S.~Yokoyama,
\newblock EPL {\bf 132}, 3 (2020), arXiv:2008.00628.

\bibitem{Mikura:2021ldx}
Y.~Mikura, Y.~Tada, and S.~Yokoyama,
\newblock (2021), arXiv:2103.13045.

\bibitem{Gunji:2019wtk}
Y.~Gunji and K.~Ishiwata,
\newblock JHEP {\bf 09}, 065 (2019), arXiv:1906.04530.

\bibitem{Adamson:2013whj}
MINOS, P.~Adamson {\em et~al.},
\newblock Phys. Rev. Lett. {\bf 110}, 251801 (2013), arXiv:1304.6335.

\bibitem{Adamson:2013ue}
MINOS, P.~Adamson {\em et~al.},
\newblock Phys. Rev. Lett. {\bf 110}, 171801 (2013), arXiv:1301.4581.

\bibitem{Abe:2017vif}
T2K, K.~Abe {\em et~al.},
\newblock Phys. Rev. D {\bf 96}, 092006 (2017), arXiv:1707.01048,
\newblock [Erratum: Phys.Rev.D 98, 019902 (2018)].

\bibitem{Abe:2018wpn}
T2K, K.~Abe {\em et~al.},
\newblock Phys. Rev. Lett. {\bf 121}, 171802 (2018), arXiv:1807.07891.

\bibitem{Adamson:2017gxd}
NOvA, P.~Adamson {\em et~al.},
\newblock Phys. Rev. Lett. {\bf 118}, 231801 (2017), arXiv:1703.03328.

\bibitem{NOvA:2018gge}
NOvA, M.~A. Acero {\em et~al.},
\newblock Phys. Rev. D {\bf 98}, 032012 (2018), arXiv:1806.00096.

\bibitem{Vagnozzi:2017ovm}
S.~Vagnozzi {\em et~al.},
\newblock Phys. Rev. D {\bf 96}, 123503 (2017), arXiv:1701.08172.

\bibitem{Ma:2001dn}
E.~Ma and G.~Rajasekaran,
\newblock Phys. Rev. D {\bf 64}, 113012 (2001), arXiv:hep-ph/0106291.

\bibitem{Babu:2002dz}
K.~S. Babu, E.~Ma, and J.~W.~F. Valle,
\newblock Phys. Lett. B {\bf 552}, 207 (2003), arXiv:hep-ph/0206292.

\bibitem{Altarelli:2005yp}
G.~Altarelli and F.~Feruglio,
\newblock Nucl. Phys. B {\bf 720}, 64 (2005), arXiv:hep-ph/0504165.

\bibitem{Altarelli:2010gt}
G.~Altarelli and F.~Feruglio,
\newblock Rev. Mod. Phys. {\bf 82}, 2701 (2010), arXiv:1002.0211.

\bibitem{Ishimori:2010au}
H.~Ishimori {\em et~al.},
\newblock Prog. Theor. Phys. Suppl. {\bf 183}, 1 (2010), arXiv:1003.3552.

\bibitem{King:2013eh}
S.~F. King and C.~Luhn,
\newblock Rept. Prog. Phys. {\bf 76}, 056201 (2013), arXiv:1301.1340.

\bibitem{King:2014nza}
S.~F. King, A.~Merle, S.~Morisi, Y.~Shimizu, and M.~Tanimoto,
\newblock New J. Phys. {\bf 16}, 045018 (2014), arXiv:1402.4271.

\bibitem{Ferrara:2010in}
S.~Ferrara, R.~Kallosh, A.~Linde, A.~Marrani, and A.~Van~Proeyen,
\newblock Phys. Rev. D {\bf 83}, 025008 (2011), arXiv:1008.2942.

\bibitem{Ferrara:2010yw}
S.~Ferrara, R.~Kallosh, A.~Linde, A.~Marrani, and A.~Van~Proeyen,
\newblock Phys. Rev. D {\bf 82}, 045003 (2010), arXiv:1004.0712.

\bibitem{Buchmuller:2012ex}
W.~Buchm\"{u}ller, V.~Domcke, and K.~Schmitz,
\newblock JCAP {\bf 04}, 019 (2013), arXiv:1210.4105.

\bibitem{Ishiwata:2018dxg}
K.~Ishiwata,
\newblock Phys. Lett. B {\bf 782}, 367 (2018), arXiv:1803.08274.

\bibitem{Coleman:1973jx}
S.~R. Coleman and E.~J. Weinberg,
\newblock Phys. Rev. D {\bf 7}, 1888 (1973).

\bibitem{Asaka:2001ez}
T.~Asaka, W.~Buchmuller, and L.~Covi,
\newblock Phys. Lett. B {\bf 510}, 271 (2001), arXiv:hep-ph/0104037.

\bibitem{Achucarro:2010da}
A.~Achucarro, J.-O. Gong, S.~Hardeman, G.~A. Palma, and S.~P. Patil,
\newblock JCAP {\bf 01}, 030 (2011), arXiv:1010.3693.

\bibitem{Chen:2009zp}
X.~Chen and Y.~Wang,
\newblock JCAP {\bf 04}, 027 (2010), arXiv:0911.3380.

\bibitem{Domcke:2014zqa}
V.~Domcke, K.~Schmitz, and T.~T. Yanagida,
\newblock Nucl. Phys. B {\bf 891}, 230 (2015), arXiv:1410.4641.

\bibitem{Domcke:2017xvu}
V.~Domcke and K.~Schmitz,
\newblock Phys. Rev. D {\bf 95}, 075020 (2017), arXiv:1702.02173.

\bibitem{Domcke:2017rzu}
V.~Domcke and K.~Schmitz,
\newblock Phys. Rev. D {\bf 97}, 115025 (2018), arXiv:1712.08121.

\bibitem{Feruglio:2017spp}
F.~Feruglio,
\newblock (2019), arXiv:1706.08749.

\bibitem{Kobayashi:2018vbk}
T.~Kobayashi, K.~Tanaka, and T.~H. Tatsuishi,
\newblock Phys. Rev. D {\bf 98}, 016004 (2018), arXiv:1803.10391.

\bibitem{Criado:2018thu}
J.~C. Criado and F.~Feruglio,
\newblock SciPost Phys. {\bf 5}, 042 (2018), arXiv:1807.01125.

\bibitem{Kobayashi:2018scp}
T.~Kobayashi {\em et~al.},
\newblock JHEP {\bf 11}, 196 (2018), arXiv:1808.03012.

\bibitem{Novichkov:2018yse}
P.~P. Novichkov, S.~T. Petcov, and M.~Tanimoto,
\newblock Phys. Lett. B {\bf 793}, 247 (2019), arXiv:1812.11289.

\bibitem{Nomura:2019jxj}
T.~Nomura and H.~Okada,
\newblock Phys. Lett. B {\bf 797}, 134799 (2019), arXiv:1904.03937.

\bibitem{Kobayashi:2019mna}
T.~Kobayashi, Y.~Shimizu, K.~Takagi, M.~Tanimoto, and T.~H. Tatsuishi,
\newblock JHEP {\bf 02}, 097 (2020), arXiv:1907.09141.

\bibitem{Ding:2019zxk}
G.-J. Ding, S.~F. King, and X.-G. Liu,
\newblock JHEP {\bf 09}, 074 (2019), arXiv:1907.11714.

\bibitem{Kobayashi:2019xvz}
T.~Kobayashi, Y.~Shimizu, K.~Takagi, M.~Tanimoto, and T.~H. Tatsuishi,
\newblock Phys. Rev. D {\bf 100}, 115045 (2019), arXiv:1909.05139,
\newblock [Erratum: Phys.Rev.D 101, 039904 (2020)].

\bibitem{Kobayashi:2019gtp}
T.~Kobayashi, T.~Nomura, and T.~Shimomura,
\newblock Phys. Rev. D {\bf 102}, 035019 (2020), arXiv:1912.00637.

\bibitem{Penedo:2018nmg}
J.~T. Penedo and S.~T. Petcov,
\newblock Nucl. Phys. B {\bf 939}, 292 (2019), arXiv:1806.11040.

\bibitem{Novichkov:2018ovf}
P.~P. Novichkov, J.~T. Penedo, S.~T. Petcov, and A.~V. Titov,
\newblock JHEP {\bf 04}, 005 (2019), arXiv:1811.04933.

\bibitem{King:2019vhv}
S.~F. King and Y.-L. Zhou,
\newblock Phys. Rev. D {\bf 101}, 015001 (2020), arXiv:1908.02770.

\bibitem{Novichkov:2018nkm}
P.~P. Novichkov, J.~T. Penedo, S.~T. Petcov, and A.~V. Titov,
\newblock JHEP {\bf 04}, 174 (2019), arXiv:1812.02158.

\bibitem{Ding:2019xna}
G.-J. Ding, S.~F. King, and X.-G. Liu,
\newblock Phys. Rev. D {\bf 100}, 115005 (2019), arXiv:1903.12588.

\bibitem{Asaka:2019vev}
T.~Asaka, Y.~Heo, T.~H. Tatsuishi, and T.~Yoshida,
\newblock JHEP {\bf 01}, 144 (2020), arXiv:1909.06520.

\bibitem{Chen:2019ewa}
M.-C. Chen, S.~Ramos-S\'anchez, and M.~Ratz,
\newblock Phys. Lett. B {\bf 801}, 135153 (2020), arXiv:1909.06910.

\bibitem{Kaplunovsky:1995jw}
V.~Kaplunovsky and J.~Louis,
\newblock Nucl. Phys. B {\bf 444}, 191 (1995), arXiv:hep-th/9502077.

\bibitem{Antoniadis:1994hg}
I.~Antoniadis, E.~Gava, K.~S. Narain, and T.~R. Taylor,
\newblock Nucl. Phys. B {\bf 432}, 187 (1994), arXiv:hep-th/9405024.

\bibitem{Asaka:2020tmo}
T.~Asaka, Y.~Heo, and T.~Yoshida,
\newblock Phys. Lett. B {\bf 811}, 135956 (2020), arXiv:2009.12120.

\bibitem{work-in-progress}
{Work in progress by Y.~Gunji, K.~Ishiwata, and T.~Yoshida}.

\end{thebibliography}
\bibliographystyle{h-physrev5}
\end{document}